\documentclass[fleqn,usenatbib]{mnras}

\usepackage{newtxtext,newtxmath}

\usepackage[T1]{fontenc}
\usepackage{ae,aecompl}

\usepackage{graphicx}	% Including figure files
\usepackage{amsmath}	% Advanced maths commands
\usepackage{amssymb}	% Extra maths symbols
\usepackage{multirow}
\usepackage{comment}
\title[Accretion of asteroids onto white dwarfs]{Accretion of tidally disrupted asteroids onto white dwarfs: direct accretion versus disk processing}

\author[Li, Mustill \& Davies]{
Daohai Li$^{1,2}$\thanks{E-mail: lidaohai@bnu.edu.cn, lidaohai@gmail.com (DL)},
Alexander J. Mustill$^1$,
and Melvyn B. Davies$^{1,3}$
\\
$^1$Lund Observatory, Department of Astronomy and Theoretical Physics, Lund University, Box 43, SE-221 00 Lund, Sweden\\
$^2$Department of Astronomy, Beijing Normal University, No.19, Xinjiekouwai St, Haidian District, Beijing, 100875, P.R.China\\
$^3$Centre for Mathematical Sciences, Lund University, Box 118, 221 00 Lund, Sweden
}

\date{Accepted XXX. Received YYY; in original form ZZZ}

\pubyear{2015}

\begin{document}
\label{firstpage}
\pagerange{\pageref{firstpage}--\pageref{lastpage}}
\maketitle

\begin{abstract}
Atmospheric heavy elements have been observed in more than a quarter of white dwarfs (WDs) at different cooling ages, indicating ongoing accretion of asteroidal material, whilst only a few per cent of the WDs possess a dust disk, and all these WDs are accreting metals. Here, assuming that a rubble-pile asteroid is scattered inside a WD's Roche lobe by a planet, we study its tidal disruption and the long-term evolution of the resulting fragments. We find that after a few pericentric passages, the asteroid is shredded into its constituent particles, forming a flat, thin ring. On a timescale of Myr, tens of per cent of the particles are scattered onto the WD, and are therefore directly accreted without first passing through a circularised close-in disk. Fragment mutual collisions are most effective for coplanar fragments, and are thus only important in $10^3-10^4$ yr before the orbital coplanarity is broken by the planet. We show that for a rubble pile asteroid with a size frequency distribution of the component particles following that of the near earth objects, it has to be roughly at least 10 km in radius such that enough fragments are generated and $\ge10\%$ of its mass is lost to mutual collisions. At relative velocities of tens of km/s, such collisions grind down the tidal fragments into smaller and smaller dust grains. The WD radiation forces may shrink those grains' orbits, forming a dust disk. Tidal disruption of a monolithic asteroid creates large km-size fragments, and only parent bodies $\ge100$ km are able to generate enough fragments for mutual collisions to be significant. Hence, those large asteroids experience a disk phase before being accreted.
\end{abstract}


%%%%%%%%%%%%%%%%%%%%%%%%%%%%%%%%%%%%%%%%%%%%%%%%%%

%%%%%%%%%%%%%%%%% BODY OF PAPER %%%%%%%%%%%%%%%%%%

\section{Introduction}\label{sec-intro}
A quarter to half of white dwarfs (WDs) show metal lines in their spectrum \citep{Zuckerman2003,Koester2014,Wilson2019}, observed at various WD cooling ages from a few times 100 Myr to several Gyr. Depending on the WD spectral type, heavy elements sink in the WD atmosphere on timescales of days to Myr \citep[e.g.,][]{Koester2009,Wyatt2014,Jura2014}. In any case, the sinking timescale is much shorter than the cooling age, implying that accretion of heavy elements onto WDs must be ongoing.

Tens of elements have been detected in the WDs' atmospheres and the overall composition is similar to terrestrial planets \citep[e.g.,][but volatile-rich pollutants have also been detected \citealt{Farihi2013}]{Farihi2009}. A widely-accepted mechanism for delivering such material to the WD is the accretion of asteroids \citep{Jura2003}. During the host star's asymptotic giant branch phase, the central star probably has cleared all material up to a few au \citep{Mustill2012}. Thus, a mechanism is needed to deliver asteroids from larger heliocentric distances to the WD surface.

A promising candidate is the interaction with planets. A planet may scatter small bodies inward or outward. Generally speaking, scatterings with giant planets are strong and rapid ejection results. In contrast, less massive (multiple) planets, especially those on eccentric orbits, have a higher chance to scatter asteroids inward towards the central host and over longer timescales \citep{Bonsor2011,Frewen2014,Mustill2018,Veras2021}. As the central host star loses its mass during the giant branch, the interplanetary forcing relative to the heliocentric gravity becomes stronger, potentially leading to the system's instability and the planets' mutual scattering \citep{Debes2002,Veras2013,Veras2016a,Maldonado2020,Maldonado2020a,Maldonado2021}. These much-excited planets may then scatter the planetesimals to the WD \citep{Mustill2018}. In the meantime, a non-negligible fraction of the planets may themselves be accreted by the WD \citep[e.g.,][]{Maldonado2021}.

In a perhaps less violent manner, planets may also send material to the WD through resonances. As the relative mass of the planet with respect to the central star increases owing to stellar mass loss, the width of the mean motion resonances becomes larger. Then objects previously not affected by resonances during the main sequence may become vulnerable in the WD phase  \citep{Bonsor2011,Debes2012,Frewen2014,Smallwood2021}. Alternatively, ingestion of close-in planets by the host star may change the dynamical structure of the system, causing the secular resonances to shift and asteroids at locations swept by the resonances may end up hitting the star \citep{Smallwood2018,Smallwood2021}.

WDs in binaries show a similar occurrence rate of atmospheric metal signs to single WDs \citep[for instance][]{Zuckerman2014,Veras2018,Wilson2019}. A companion star allows for other avenues for WD pollution or enhances ones operating for single stars already. For relatively close binaries, during the stellar mass loss, the planet's orbit expands more rapidly compared to the stellar binary trajectory and may become unstable if it becomes too close to the companion \citep{Kratter2012}. The differential orbital expansion may enhance the Kozai-Lidov (KL) cycles, forcing objects onto the WD \citep{Stephan2017}.  Alternatively, engulfment of close-in planets may leave the previously-protected asteroids subject to strong KL effects \citep{Petrovich2017}. Finally, the companion orbit may itself become highly eccentric due to galactic tide on the timescale of Gyr, destabilising the planetary systems \citep{Bonsor2015}.

While accretion of asteroidal material seems rather ubiquitous, a far smaller fraction of the WDs, maybe a few per cent, possess dust disks within the stellar Roche radius, betrayed by an infrared excess \citep[e.g.,][]{Farihi2009,Wilson2019}. The disk is probably geometrically like the Saturnian rings \citep{Jura2003,Rafikov2011}, formed through circularisation of tidal fragments of a large object inside the Roche lobe of the WD \citep{Debes2012,Veras2014,Veras2015a,Malamud2020,Malamud2020a}.

Several authors have looked into the tidal disruption of asteroid/planets around WDs in isolation \citep{Debes2012,Veras2014,Malamud2020,Malamud2020a}, without considering the existence of another planet that throws the object close to the WD in the first place. In this work, we bridge the gap and study the tidal disruption of an asteroid and the ensuing long term evolution of the tidal fragments under the effect of a planet, radiation forces and mutual collisions.

We organise the paper as follows. In Section \ref{sec-tidal}, we model the tidal disruption of asteroids close to a WD and show how a ring of fragments results. Then in Section \ref{sec-planet}, the interaction between the tidal disruption fragments and the planets is under investigation. Section \ref{sec-coll} is devoted to the collisional evolution among the fragments. The implications are discussed in Section \ref{sec-dis}. Finally, we present the conclusions in Section \ref{sec-con}.

Before presenting the details, we first clarify the terms used in this work. We are concerned about the tidal disruption of small bodies around a WD and the following evolution. For ease of reference, a small body, before its disruption, will be called an {\it asteroid}. After disruption, the remnants shredded from the asteroid will be referred to as {\it fragments}. Both asteroids and fragments, as seen below, will be modelled as ensembles of small particles. So {\it particles} will be used only to refer to the components of the asteroid.

\section{Tidal disruption of small bodies near a WD}\label{sec-tidal}
When an asteroid comes inside the Roche of the central WD, it may be tidally disrupted \citep[for instance][]{Debes2012,Veras2014}. In this section, we numerically model this process.

\subsection{Methods}

A number of works have looked into the disruption of an asteroid around a more massive object using numerical methods \citep[see][for early and recent examples]{Asphaug1996,Zhang2020}. Here we model the asteroid as a non-cohesive rubble pile \citep[see e.g., ][]{Richardson1998} and follow its evolution in the gravitational field of a WD. Our code couples granular dynamics with Newtonian gravity\footnote{The $N$-body part of our code is adapted from a program by Raymond J. Spiteri, as in the lecture notes of Parallel Programming for Scientific Computing \url{https://www.cs.usask.ca/~spiteri/cmpt851.html}.} as we describe below.

An asteroid is modelled as a collection of constituent particles. Each particle is represented by a hard sphere\footnote{A soft-sphere approach will not change the qualitative conclusions \citep[e.g.][]{Zhang2020}.} of its own mass, radius, position, velocity and spin. For simplicity, in the same simulation, all particles have the same physical properties. The particles' velocity and spin are affected by their mutual gravity and collisions and a kick-drift-kick leapfrog scheme is used to propagate these state vectors. In a kick, our code uses a brute-force method to calculate the gravity. The evaluation of gravity is not the main factor limiting the performance, as this is done only once in an entire timestep. Also in a kick, a list of neighbours for each particle is constructed. In a drift step, we first search for collisions in the neighbour list for each particle that would happen during the drift in accordance with \citet{Richardson2009}; these are subsequently sorted by their collision time. Then if needed, sub-steps are performed so that the particles involved in the earliest collision are just touching. Their relative position and velocity at the contact point are calculated and the post-collision velocity and spin are derived using the coefficients of restitution in the normal and tangential directions as per \citet{Richardson1994,Richardson1995,Richardson1998}. Now, a new search for collisions is done for those two particles only. We repeat this process until all collisions above-registered are dealt with. Then the system is drifted to the point where the next kick is due.

When processing the output containing the state vectors of each particle, a friend-of-friend method is employed to look for aggregates of particles (fragments). Here, we have used a rather small threshold such that the particles of the same fragment are more or less touching. We note that binary fragments, if any, are merged if their mutual distance is smaller than the Hill radius with respect to the central host.

\subsection{A case study for a pericentric distance of $4\times10^{-3}$ au}\label{sec-004}

\begin{figure}
\includegraphics[width=\columnwidth]{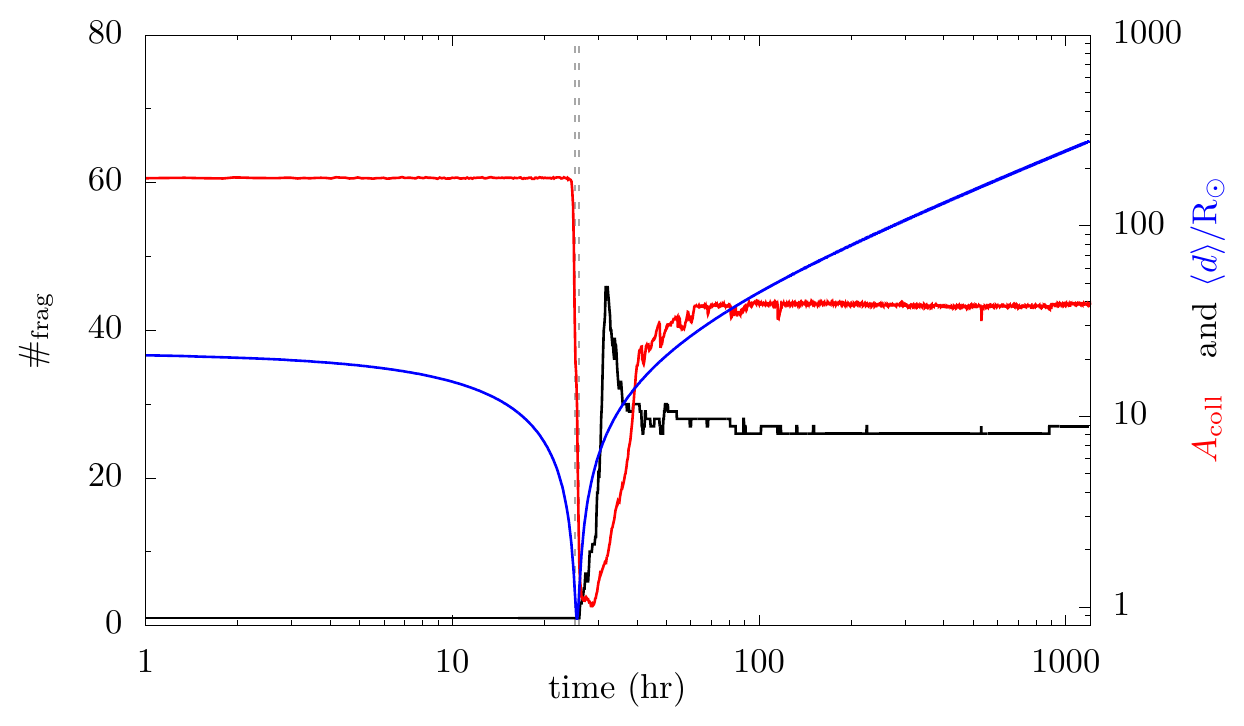}
\caption{Tidal disruption of an asteroid near a WD. On the left ordinate, we show the time evolution of the number of fragments $\#_\mathrm{frag}$ in black, and on the right $y$-axis the relative inter-particle collision intensity $A_\mathrm{coll}$ in red and the distance from the centre of mass of all particles to the central host $\langle d \rangle$ in blue. $A_\mathrm{coll}$ has been normalised with respect to its lowest value during the evolution and $\langle d \rangle$ against the solar radius. $\langle d \rangle$ is measured for all particles, so representing all fragments after the disruption. The two vertical grey lines mark the moments when $\langle d \rangle$ is equal to the nominal tidal disruption distance of 1.24 $\mathrm{R}_\odot$.}
\label{fig-time-evo-20}
\end{figure}

First, we perform test simulations to examine the performance/convergence of the code. The mass of the central WD is 0.75 $\mathrm{M}_\odot$ (solar mass), as appropriate for a progenitor star of about 3 $\mathrm{M}_\odot$. Our three model asteroids are composed of 1000, 4000, or 8000 particles respectively, each of 1, 0.63 and 0.5 km in radius (the sums of the volumes of all individual particles in all three resolutions are the same) with the same particle density of 2.7 g cm$^{-3}$. We assemble the asteroid by letting the particles free fall from randomised positions. In all three resolutions, the resulting asteroid radius is about 11.6 km, implying a bulk density of 1.74 g cm$^{-3}$ and a porosity of 0.35. The nominal tidal disruption distance for these asteroids, according to \citet{Sridhar1992,Rafikov2018}, is then 1.24 solar radii (about $6\times10^{-3}$ au). Here for our test simulations, the pericentre ($q$) and apocentre distances of the asteroids are $4\times10^{-3}$ and 10 au, respectively and the initial heliocentric distance is 0.1 au. The simulation is run for 1200 hr (50 days), far more than enough for the reaccumulation process after disruption, if any \citep[][and see Figure \ref{fig-time-evo-20}]{Hahn1998,Malamud2020}. The exact restitution coefficients seem not important as long as the collision is inelastic \citep{Richardson1998}. For the resolution of 1000 particles, we have run 20 simulations with different random number seeds; among these, 10 are with the two restitution coefficients being 0.8 and the other 10 with 0.55 \citep{Richardson1998,Zhang2020}. The resolution of 4000 and 8000 are computationally more expensive and we have only one run for each, both with coefficients of restitution 0.8. Our timesteps for the three resolutions are 3.8, 2.4, and 1.9 sec such that it takes tens of timesteps for a particle, moving at escape velocity of the asteroid, to pass by another particle and collisions can be sufficiently resolved. In the following, we refer to the asteroid of 1000 particles, radius of 11.6 km and restitution coefficients of 0.8 as our nominal asteroid setup.

\begin{figure}
\includegraphics[width=\columnwidth]{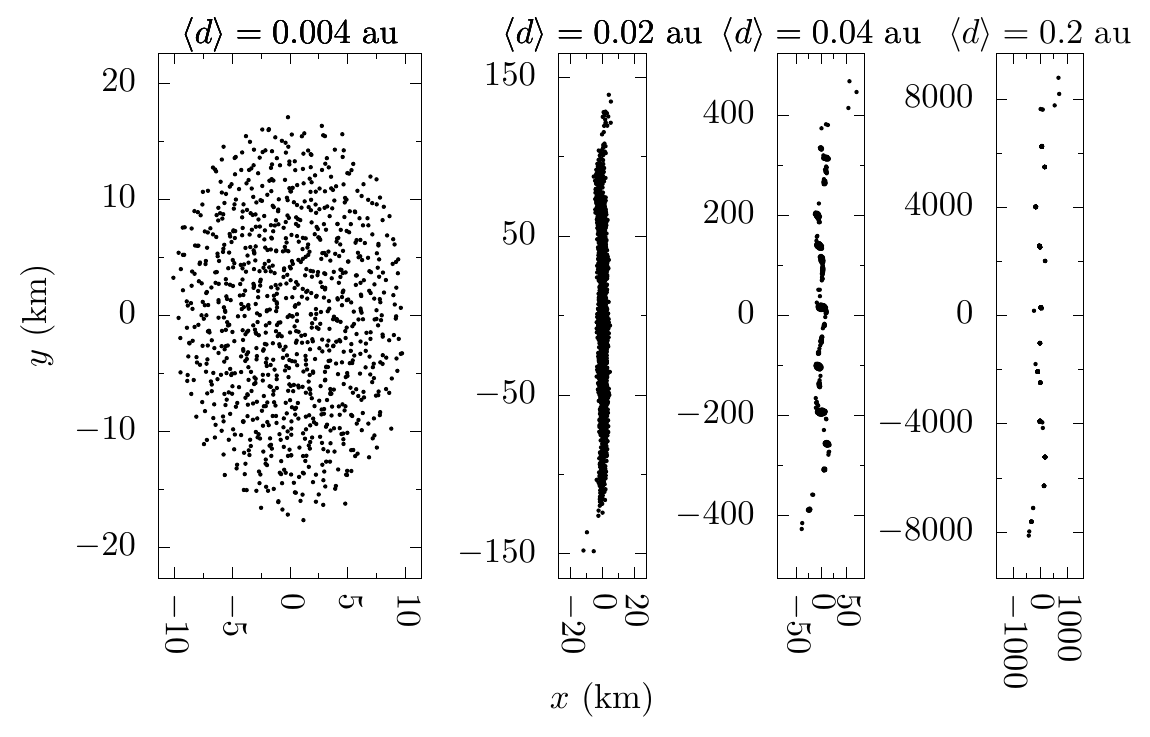}
\caption{Snapshots of the tidal disruption of a rubble pile asteroid near a WD. The panels show the positions of the asteroid's composing particles projected onto the orbital plane at different times. The distance between the WD and the centre of mass of the asteroid is marked on top of each panel. The positions of the particles have been rotated for better presentation and the $x$- and $y$- axes have the same scale within each panel.}
\label{fig-snapshot}
\end{figure}

In Figure \ref{fig-snapshot}, we show the snapshots of the disruption of the nominal asteroid. The left panel shows the moment when the asteroid is at its pericentre $4\times10^{-3}$ au. Now, though apparently deformed, the component particles still stay together. Soon at $2\times10^{-2}$ au, the asteroid becomes a string of particles under the tidal field of the WD. Moving further, the string collapses under self-gravity and a few tens of fragments form.

Figure \ref{fig-time-evo-20} shows the time evolution of the number of fragments $\#_\mathrm{frag}$ (black, left $y$-axis), the relative inter-particle collision intensity for all fragments $A_\mathrm{coll}$ (red, right $y$-axis) and the distance between the centre of mass of all particles and the WD $\langle d \rangle$ (blue, right $y$-axis) from a simulation with the nominal asteroid setup. The collision intensity records the number of collisions counted in a given time interval; this has been normalised by the lowest value registered in the simulation. As can be seen, $A_\mathrm{coll}$ remains constant when the asteroid is far from the WD, as the structure of the asteroid is not changed. When the asteroid is close to the Roche sphere of the WD (the two vertical grey lines mark the enter and exit), $A_\mathrm{coll}$ starts and continues to drop, meaning that the asteroid is being shredded and the particles are drifting away from each other. But at this moment, the number of fragments $\#_\mathrm{frag}$ is still 1, so the asteroid's component particles are still pretty much touching (though not colliding with) each other. After leaving the Roche sphere, particles continue to separate under Kepler shear and $\#_\mathrm{frag}$ increases. But finally, particles reaccumulate to form larger fragments under self-gravity so $\#_\mathrm{frag}$ decreases and $A_\mathrm{coll}$ increases.% The reaccretion process has to compete with Kepler shear which tend to further stretch the stream \citep{Hahn1998}.

\begin{figure}
\includegraphics[width=\columnwidth]{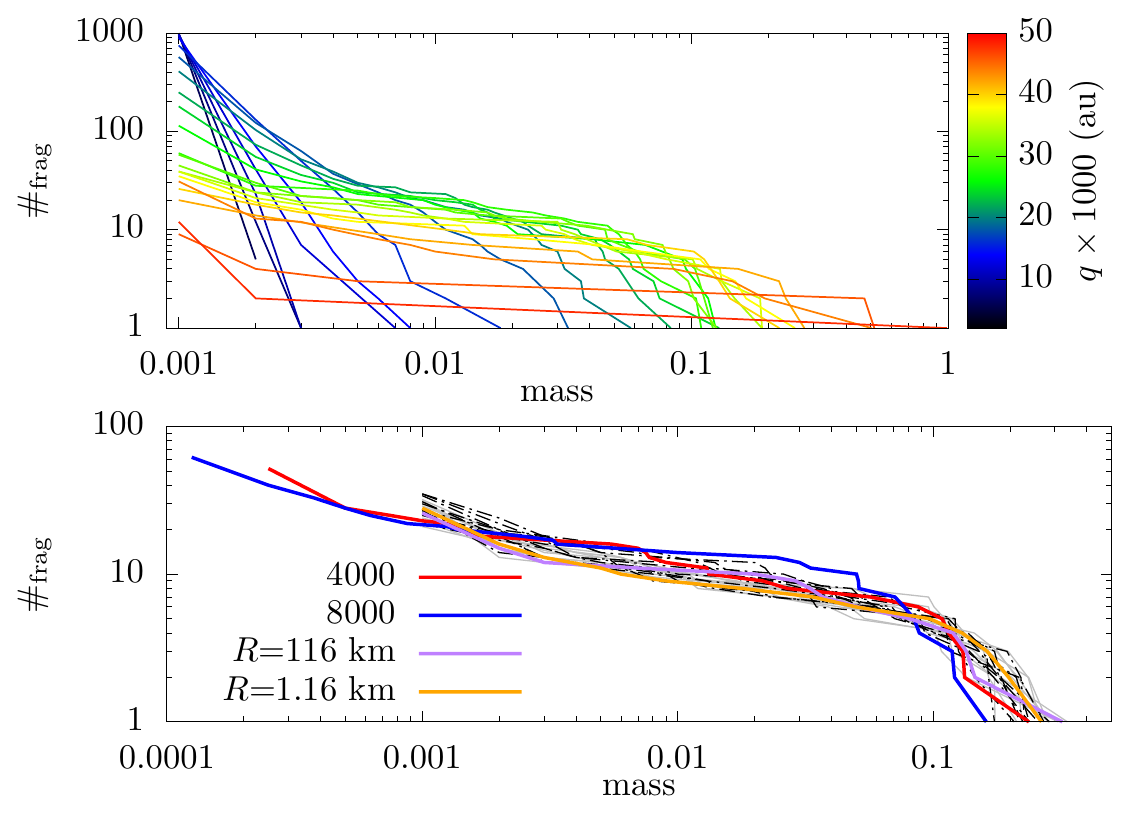}
\caption{Size frequency distribution (SFD) of the tidal fragments from an asteroid. The $x$-axis is the fragment mass fraction with respect to the parent asteroid and $y$-axis shows the number of fragments with masses no smaller than the value marked on the $x$-axis. Top: the SFDs for different pericentric distances as shown in different colours, all parent asteroids with the nominal setup. Bottom: SFDs for different setups for the asteroids, all at a fixed pericentre distance of $4\times10^{-3}$ au. The grey solid and black dash-dotted lines, each containing 10 instances of different random number seeds, are with restitution coefficients 0.8 and 0.55, respectively. The red and blue lines present simulations with asteroids resolutions of 4000 and 8000 particles. The purple line show the result of the disruption of a parent asteroid that is 116 km in radius; for the orange line, the asteroid radius is 11.6 km.}
\label{fig-converg}
\end{figure}

The reaccreted fragments follow a distribution in their physical sizes and orbits. In the bottom panel of Figure \ref{fig-converg}, we show the size frequency distribution (SFD) of the above (20+2) simulations. The grey and black lines present the results with the asteroid resolution of 1000 particles, for restitution coefficients 0.8 and 0.55, respectively. The red and blue lines are for simulations with resolutions of 4000 and 8000 particles. Generally speaking, all these simulations of different resolutions/restitution coefficients/random number seeds are broadly similar though with a spread of a factor a few. It seems that the SFD can be approximated by a broken power law, perhaps visually not so different from that of \citet{Debes2012,Malamud2020a}, steeper at larger sizes and shallower at smaller sizes. The distribution of fragments from tidal disruption events have been looked into in the context of a near earth asteroid flying-by the Earth \citep{Schunova2014}. There, the SFDs were derived for the NEA population averaged for different Earth-encounter scenarios and a single power law applied.

In order to test the effect of the asteroid's size, we here perform two additional simulations. The resolution is kept at 1000 particles but the asteroid radius set to 116 km and 1.16 km, respectively. The SFD of the fragments from these two simulations are shown as the purple and orange lines in the bottom panel of Figure \ref{fig-converg}, both consistent with previous simulations where the asteroids are 11.6 km. The reason may be that the reaccumulation is controlled by the competition between the stretch of the fragment stream out of the tidal disruption owing to the Keplerian shear and local free fall \citep{Hahn1998}. A larger asteroid radius leads to a wider spread in the post-disruption orbits (cf Equation \eqref{eq-a-e-r}), increasing the timescales of both phenomena, so perhaps these effects somehow cancel out.

\begin{figure}
\includegraphics[width=\columnwidth]{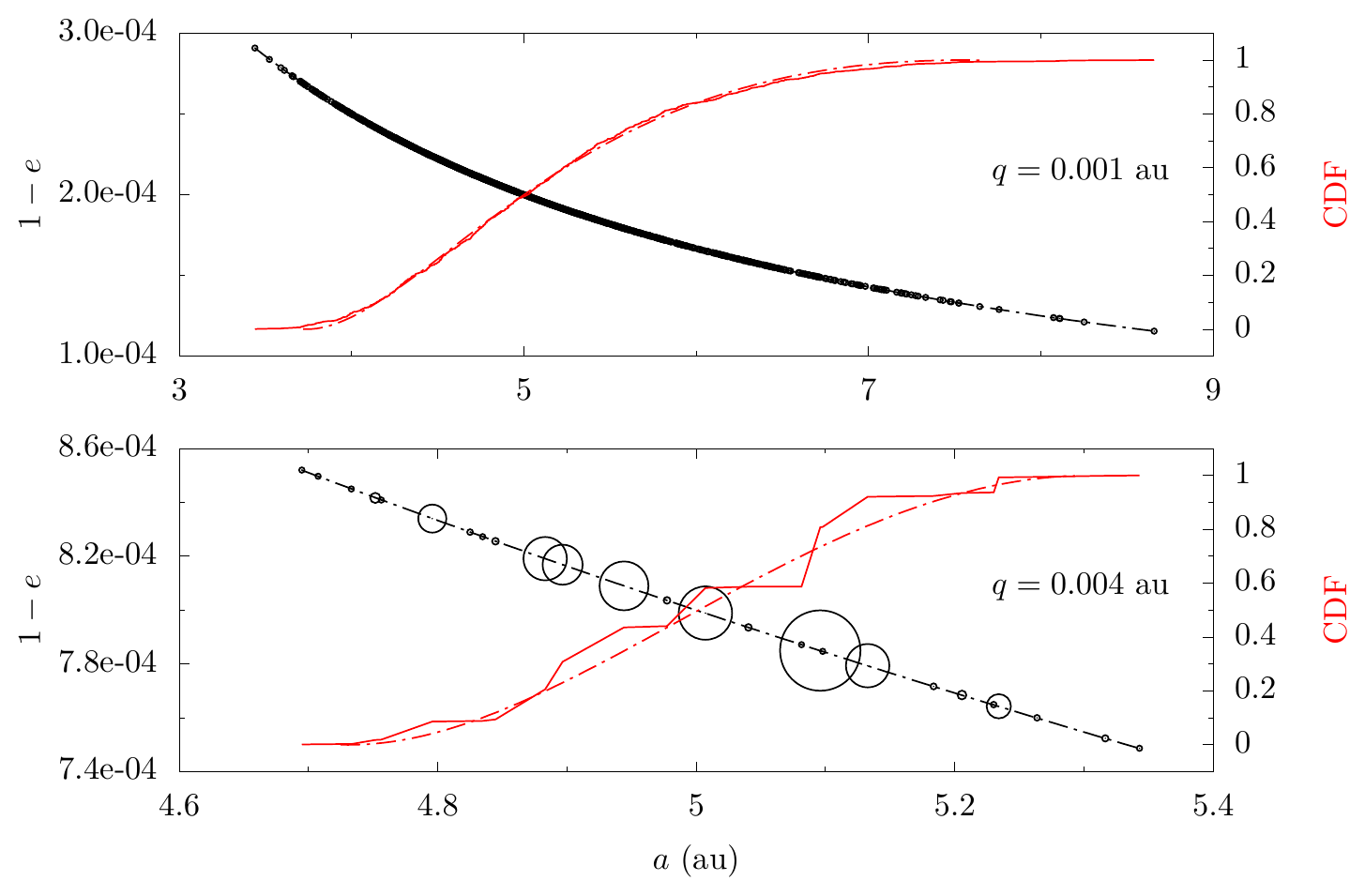}
\caption{Orbital distribution for the fragments from tidal disruption events for $q=1\times10^{-3}$ (up) and $4\times10^{-3}$ au (bottom). The $x$-axis is semimajor axis $a$. The left ordinate is the fragment's $e$ shown in black. The point size is related to the physical size of a fragment. The right $y$-axis is the cumulative distribution function (CDF) of $a$ in red. The black and red dash-dotted lines represent that expected from Equation \eqref{eq-a-e-r} with $r_0$ fitted from the simulations.}
\label{fig-a-e-size-20}
\end{figure}

The distribution of the fragments' semimajor axis $a$ and eccentricity $e$ are shown in the bottom panel of Figure \ref{fig-a-e-size-20} for a simulation randomly picked from the ten of the nominal asteroid setup. The black circles present $e$ of the observed fragments, radii proportional to the physical sizes and corresponding to the left $y$-axis. The two quantities are obviously correlated. The cumulative distribution function (CDF) of $a$ is shown on the right $y$-axis in red. Here the CDF is counted such that the contribution from a fragment is proportional to its mass, the reason why there are discontinuous jumps in the plot.

Now we follow the impulse approximation \citep[e.g.,][]{Veras2014} to understand the result. The assumption is that the mutual gravity between the constituent particles of an asteroid becomes instantly negligible upon the tidal disruption and the fragments/particles then move on Keplerian orbits. On disruption, all particles have the same heliocentric velocity and hence the same specific kinetic energy. Any resultant difference between the orbits of different particles/fragments must come from the difference in their position at that moment. Suppose we have a particle/fragment and the differences in the specific energy $E$ and angular momentum $L$ between it and those of the centre of mass of the asteroid upon disruption, are
\begin{equation}
\label{eq-a-e-r}
\begin{aligned}
\delta r&=r-r_0,\\
\delta E&=-{GM_*\over r}+{GM_*\over r_\mathrm{0}}\sim -GM_* {\delta r/r^2_0}\approx -{GM_*\over 2a}+{GM_*\over 2a_0},\\
\delta L&= r\,v_\mathrm{0,r}-r_\mathrm{0}\,v_\mathrm{0,r}=\delta r v_\mathrm{0,r}\approx\sqrt{GM_*a(1-e^2)}-\sqrt{GM_*a_0(1-e_0^2)}.
\end{aligned}
\end{equation}
Here $G$ is the gravitational constant, $M_*$ the mass of the central star, $r$ the heliocentric distance of the particle/fragment on disruption, $v_\mathrm{r}$ the velocity component perpendicular to the position vector on disruption and $a$ and $e$ its orbital semimajor axis and orbital eccentricity post disruption; quantities with subscript 0 represent those of the centre of mass of the asteroid. On the right hand side of the equations, the orbital energy and angular momentum have been expressed in orbital elements \citep{Murray1999}. Also, because an asteroid's physical radius is much smaller than its orbital pericentric distance $\delta r<R\ll r$, we have expanded the left hand side of the equation using $\delta r/r$. From the above equation, apparently both $a$ and $e$ of a particle are uniquely defined by its $\delta r$, indicating a link between $a$ and $e$.

A closer inspection of Figure \ref{fig-a-e-size-20} suggests that the resulting fragments do not overlap with each other in the orbital space. This can be explained by the conservation of the asteroid volume before and after disruption \citep[cf.][]{Tanga1999}. Assume that the composing particles of the same fragment are also close together in the parent asteroid before the disruption characterised by its $\delta r$. Then large fragments cannot have the same $\delta r$ owing to the finite volume of the asteroid. Therefore, a fragment has a unique $\delta r$ which according to Equation \eqref{eq-a-e-r} determines its orbital elements post disruption.

We have yet to determine the equivalent disruption distance $r_0$. This $r_0$ is not necessarily the pericentric distance \citep{Malamud2020}, nor has it to be the Roche radius \citep{Sridhar1992,Steinberg2019} so we fit it using the distribution of $a$ of the fragments. A complication is that the asteroid radius $R$ also plays a role. At the time of disruption, the asteroid may be deformed already and its $R$ may not be the original value. But we simply assume that $R$ is a constant, equal to the initial value. When $r_0$ is obtained, we can plot the CDF for $a$ and the $a-e$ relation as advised by Equation \eqref{eq-a-e-r}, shown as dash-dotted lines in Figure \ref{fig-a-e-size-20}. The agreement with data is excellent.

As a major aim of this work is to study how the disruption process affect the ingestion of asteroid by the WD, it is natural to ask if the fragments have different pericentric distances. Using $\delta r/r_0\ll 1$ and $1-e\ll1$ (the the orbits are extremely eccentric) and making some algebraic manipulations of the second line of Equation \eqref{eq-a-e-r}, it can be shown that $(\sqrt{q}-\sqrt{q_0})/\sqrt{r_0}\sim\delta r/r_0$. Therefore, fragments of the same tidal disruption event have the same pericentre distances as the parent body. Finally, the dispersion in orbital inclination is tiny, of the order of $\sim 10^{-3}$ deg in our simulations.

\subsection{Varying the pericentric distance}

Until now, we have established the SFD of the fragments, their relationship between $a$ and $e$ for the pericentre distance $q$ of $4\times10^{-3}$ au. The outcome of the tidal disruption event obviously depends on $q$ which affects both the orbital and the physical properties of the fragments \citep[e.g.,][]{Schunova2014}. Thus, we now perform a set of tidal disruption simulations with varying $q$.

The asteroids are assembled the same as in the nominal setup. The restitution coefficients in the two directions are 0.8. The pericentre is varied between $2\times10^{-4}$ and $6\times10^{-3}$ au with an increment of $2\times10^{-4}$ au, thus a total of 30 simulations.
 
As expected, a smaller pericentre leads to a wider orbital spreading and faster stretching among the particles, inhibiting reaccumulation and leading to smaller fragments. The top panel of Figure \ref{fig-converg} shows the SFD from our simulations, different colours representing different $q$. From the figure, it seems that when $q$ is small ($\lesssim 1.5\times10^{-3}$ au), the SFD is more like a single power law whereas when $q$ is large ($\gtrsim 1.5\times10^{-3}$ au), a broken power law provides a better fit.

The larger tidal field associated with a smaller $q$ also results in a larger spreading in the orbits of the fragments. An example for $q=1\times10^{-3}$ au is shown in the top panel of Figure \ref{fig-a-e-size-20} where the distribution of $a$ now ranges from 3.5 up to 8.5 au. And the fragments' $e$ spans a wider range range as well compared to $q=4\times10^{-3}$ au in the bottom panel. We have in this figure, also plotted the fitted distribution of $a$ with an $r_0$ and the agreement with the data is fairly well. For two pericentre distances $q=2\times10^{-4}$ and $4\times10^{-4}$ au, a fraction of the fragments are ejected with hyperbolic orbits. According to the analytical arguments of \citet{Rafikov2018}, to eject part of the fragments from our test asteroid, $q$ as large as $8\times10^{-4}$ au may still be possible. Here from our simulation, it seems smaller $q$ may be required.

\subsection{Repeated pericentric passages}\label{sec-rep}
After a single pericentric passage, the fragments, formed by reaccretion of the parent asteroid's composing particles, attain different sizes (Figure \ref{fig-converg}). During the ensuing close passages between them and the central star, the large fragments may be torn apart to smaller and smaller pieces.

We model this process with our ``long-term'' disruption simulations spanning many orbits. In our previous simulations, the apocentre distance of the asteroid is 10 au, implying an orbital period of more than 10 yr. To model the gravitational and collisional interaction between thousands of particles for tens of years is beyond our computational resources \citep[but the fragments may be analytically propagated when far from the WD; see][]{Malamud2020}. So here for these many-orbit simulations, we let the asteroid's apocentre be 0.4 au, the same as \citet{Veras2014}. The corresponding orbital period is about 38 days and we track the particles' evolution for 800 days. Two pericentric distances are examined, $2\times10^{-3}$ and $4\times10^{-3}$ au, both leading to partial disruption during the initial crossing of the pericentre (cf. Figure \ref{fig-converg}). The other parameters are the same as our nominal simulation.

\begin{figure}
\includegraphics[width=\columnwidth]{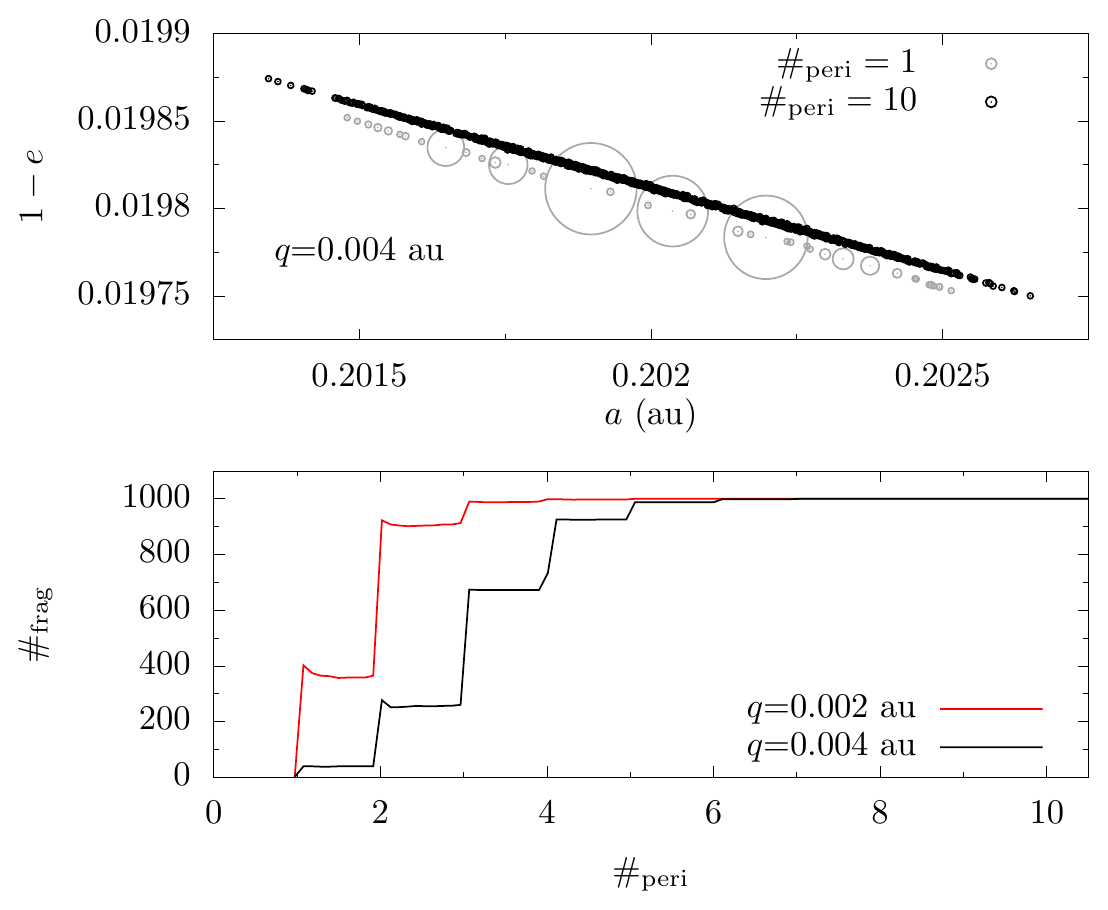}
\caption{Tidal disruption by repeated pericentric passages. The bottom panel shows the number of fragments as a function of the number of pericentric passages $\#_\mathrm{peri}$, red for $q=2\times10^{-3}$ au and black for $q=4\times10^{-3}$ au. The asteroid is initially composed of 1000 particles. The top panel show the orbital distribution of the fragments for $q=4\times10^{-3}$ au at $\#_\mathrm{peri}=1$ (grey) and at 10 (black). The ordinate in the latter has been shifted vertically by $10^{-5}$ for better visibility.}
\label{fig-a-e-size}
\end{figure}

The bottom panel of Figure \ref{fig-a-e-size} shows the number of fragments $\#_\mathrm{frag}$ as a function of pericentric passages $\#_\mathrm{peri}$ for the two simulations. Take the black line for $q=4\times10^{-3}$ au for instance. On the initial disruption $\#_\mathrm{peri}=1$, the parent asteroid is split into 40 pieces, consistent with that in Figure \ref{fig-converg}. Then upon the second passage, more than 200 fragments are created out of the previous 40. Further splitting goes on until after 6 passages, no fragment composed of two or more particle exists. The story is the same for $q=2\times10^{-3}$ au, the difference being that the final state is reached earlier owing to the smaller pericentre distance.

In the top panel of Figure \ref{fig-a-e-size}, we show the orbital distribution in $(a,e)$ of the fragments at $\#_\mathrm{peri}=1$ (grey) and those at $\#_\mathrm{peri}=10$ when totally shredded (black); the two are vertically shifted for better visibility. The profiles of the two distributions agree rather well, the only difference being that that at $\#_\mathrm{peri}=10$ is somewhat broader. The agreement can be understood by applying the above volume conservation analysis repeatedly to fragments from early disruptions.

Overall, because the size of the asteroid is small, the energy dispersion is small (Equation \eqref{eq-a-e-r}). As a consequence, the fragments, after multiple disruption events, form a flat narrow ring \citep{Veras2014}, rather than an extended disk, as in the case of the disruption of a planet \citep{Malamud2020}.

The fragments being repeatedly ground to the constituent particles suggests that the final fragment size depends on the simulation resolution. Here our asteroid is modelled as a rubble pile where only gravity is at work and material cohesion is missing. In the solar system, asteroids $\lesssim$ 10 km are thought to be such rubble piles as often indicated by an lower limit of 2 hr of the asteroids' rotation period \citep{Pravec2002}. Such small asteroids possibly come from reaccumulation of parent asteroids after a collision \citep{Walsh2018}. On the other hand, though this limit is also observed for much larger objects, gravity dominates over material strength so not much can be said on the latter \citep{Holsapple2007}. Small rubble-pile near earth asteroids have grain sizes from mm to tens of m with a power law size frequency distribution (SFD) with an exponent around 3 but the scatter is a few times 0.1 \citep{Mazrouei2014,Michikami2019,DellaGiustina2019}. While the large boulders are probably primordial from the collisional creation of the asteroids from the parent body, the smaller particles may have formed later via, e.g., collisions with small objects.

On the other hand, if the parent asteroid is monolithic, the internal material strength would counter-balance tidal fragmentation. The larger the pericentre (or the higher the strength) the larger the fragment that is resistant to further disruption \citep{Kenyon2017,Rafikov2018,Manser2019,Zhang2021}. Fragments of tens of m close to the WD surface or hundreds of kilometres close to the Roche radius, are expected.

Therefore, our simulation here suggests that rubble pile asteroids will be shredded into particles covering a large size range from mm to tens of m while monolithic asteroids will be torn into large km-size fragments. The fragment size has important implications on the collisional evolution as shown in Section \ref{sec-coll}.

%%%%%%%%%%%%%%%%%%%%%%%%%%%%%%%%%%%%%%%%%%%%%%%%%
\section{Orbital evolution of the fragments}\label{sec-planet}
In the previous section, we have shown that an asteroid, if plunging close to the WD, will be shredded down to small particles/fragments. Now we discuss the fragments' ensuing evolution.

How an asteroid gets to close to the WD in the first place to be tidally disrupted is not modelled in this work. Among the other candidates \citep[e.g.,][]{Kratter2012,Petrovich2017,Veras2020b}, one is the perturbation by the planets in the same system \citep{Bonsor2011,Debes2012,Frewen2014,Mustill2018}. After the disruption, except for some (if any) on hyperbolic orbits \citep{Rafikov2018}, the fragments' orbits continue intersecting that of the perturber. Furthermore, with pericentric distances within a solar radius, the WD radiation force may also cause significant orbital changes \citep{Veras2015a}. In this section, we model the evolution of the fragments' orbits under the perturbation of planets and radiation using the $N$-body package {\small MERCURY} \citep{Chambers1999}.

\subsection{Initial conditions}
Our model planetary system is comprised of a central WD, one or two planets and fragments. The WD properties and tidal fragments are taken from Section \ref{sec-tidal}. The assumption that the planet scatters the asteroid towards the WD implies a close encounter between the two right before the asteroid's disruption. Therefore, the orbital phases of the parent asteroid and the planet must fulfil some configuration constrained by the encounter, as we discuss below.

We first test a one-planet scenario. In all cases, the planet is revolving around the WD with a semimajor axis of $a_\mathrm{P}=10$ au. In ``q2Q10R10-e0i0mN'', the parent asteroid pericentric distance is $q_0=2\times10^{-3}$ au, apocentric distance $Q_0=10$ au and its physical radius is 11.6 km. The tidal fragments from such an asteroid after the first pericentric passage as in Section \ref{sec-tidal}, a total of $\sim400$ (Figure \ref{fig-converg}), are added to {\small MERCURY}, keeping the phase (e.g., the longitude of pericentre is $0^\circ$). Those are treated as ``small bodies'' meaning that they are not interacting with each other. The planet's orbit is circular ($e_\mathrm{P}=0$) and coplanar with respect to those of the fragments ($i_\mathrm{P}=0$), and the planet's mass is that of Neptune. In this case, encounters between the asteroid and the planet can only happen at the asteroid's apocentre, so only one phase configuration is feasible.

In the second case, we adopt $q_0=4\times10^{-3}$ au for the pericentric distance of the asteroid as this may change the frequency of accretion of fragments by the WD. However, tidal disruption at this distance only creates a few tens of fragments after the first pericentric passage (Figure \ref{fig-converg}). So we have randomly picked 14 out of the 20 simulations done for this pericentre distance as in Section \ref{sec-004}. The fragments from all these 14, totalling about 400, are scaled so the total mass is the same as the other cases (though their mass plays no role here as the fragments are treated as small bodies in {\small MERCURY}). This simulation is referred to as ``q4Q10R10-e0i0mN''.

When $Q_0=10$ au and $e_\mathrm{P}=0$ (the planet's semimajor axis is always $a_\mathrm{P}=10$ au), the asteroid's and the planet's orbits barely touch, representing a case of minimum scattering between the fragments and the planet. With a larger $Q_0$, there is substantial orbital overlap and stronger scattering may result. Here we simply let $Q_0=12$ au in our third case. Now, two phase configurations are allowed: the planet and the parent asteroid encounter when the asteroid is moving towards/away from its apocentre. This is our ``q2Q12R10-e0i0mN'' simulation.

Next, we change the asteroid radius to 116 km, which causes a wider spread in the fragments' orbits. This is our ``q2Q10R100-e0i0mN'' simulation.

Then we vary the planet parameters. In ``q2Q10R10-e6i0mN'' and ``q2Q10R10-e0i6mN'', we let its eccentricity $e_\mathrm{P}=0.6$ and inclination $i_\mathrm{P}=60^\circ$, respectively. In the former case, the parent asteroid's and the planet's orbits may interact at different locations. In order to reduce the freedom, we let the two objects meet at a heliocentric distance of 10 au. Hence, the asteroid must be then at its apocentre but the planet's eccentric orbit allows for two phase configurations, it approaching/receding from the WD. For the latter case ``q2Q10R10-e0i6mN'', also two possibilities exist: encounter at ascending/descending node. And in this case, the fragments' inclination is measured against the orbital plane of the planet, when processing the output. Then in ``q2Q10R10-e0i0mS'' the planet is assigned a Saturnian mass. Finally, ``q2Q10R10-e0i0mN+P2'' represents a two-planet scenario. The second planet's orbit has a semimajor axis of 15 au and is circular and coplanar \citep[such a configuration is long term stable, e.g.,][]{Gladman1993}. Its orbital phase is randomly chosen.

Now we move to the radiation forces which can be important for small pericentric distances. A WD's luminosity depends on its cooling age. At a cooling age of 100 Myr, the luminosity is very roughly of the order of one per cent of the current solar value ($L_\odot$) \citep{Veras2016}. Small objects with large surface area to mass ratios are the most affected and for those, we may assume a uniform temperature \citep{Broz2006}. Hence, effects related to heat latency like the Yarkovsky effect can be omitted \citep{Veras2015a}. Then only the instant absorption and reflection of the incident radiation matters, which we refer to broadly as Poynting–Robertson (PR) drag \citep{Veras2015}. It acts to shrink and circularise the fragments' orbits on timescales of thousands to millions of years depending on fragment physical size and WD luminosity \citep{Veras2015a}. For larger objects, the Yarkovsky effect may be non-negligible \citep{Veras2015}; but the overall effect is anyway small because of their lower surface area to mass ratios.

Here we run two sets of simulations, incorporating PR-drag following \citet{Veras2015,Veras2015a}. In both, the initial setup is the same as q2Q10R10-e6i0mN. But now each fragment is randomly assigned a physical size with its logarithmic radius evenly distributed between 1 mm and 1 km. In the two sets, the WD luminosity is $L=$ 0.01 $L_\odot$ and 0.001 $L_\odot$, representing a young and an old WD, respectively. The two sets are called ``q2Q10R10-e6i0mN-PRY'' and ``q2Q10R10-e6i0mN-PRO''.

\begin{table*}
\centering
\caption{Configuration of the planet-fragments interaction simulations. In all the setups, the planet that scatters the parent asteroid of the fragments to the WD has a semimajor axis of 10 au; the other parameters are varied as below. The first column lists the label of that setup. Then in columns 2, 3 and 4, the asteroid's pericentric and apocentric distances $q_0$ and $Q_0$ and its physical radius are listed. The following three list the orbital eccentricity and inclination and the mass (Nep for a Neptunian mass and Sat for Saturnian) of the planet. The eighth column shows whether another planet is included in the simulation. The last indicates whether PR-drag is included and if so the WD luminosity adopted.}
\label{tab-conf-plt}
\begin{tabular}{c c c c c c c c c c}
\hline
\multirow{2}{*}{label}&\multicolumn{3}{c}{asteroid property}&\multicolumn{3}{c}{planet1 property}&\multirow{2}{*}{planet 2}&PR\\
  & $q_0$ ($\times10^{-3}$ au)&$Q_0$ (au)& radius (km)&$e_\mathrm{P}$&$i_\mathrm{P}$ (deg)&mass&&$L_\mathrm{WD}$ ($L_\odot$)\\
\hline
q2Q10R10-e0i0mN & 2& 10& 10& 0&0&Nep&-&-\\
q4Q10R10-e0i0mN & 4&10& 10& 0&0&Nep&-&-\\
q2Q12R10-e0i0mN & 2&12& 10&0 &0&Nep&-&-\\
q2Q10R100-e0i0mN & 2& 10& 100& 0&0&Nep&-&-\\
q2Q10R10-e6i0mN & 2&10& 10&0.6& 0&Nep&-&-\\
q2Q10R10-e0i6mN & 2&10& 10&0& 60&Nep&-&-\\
q2Q10R10-e0i0mS & 2& 10& 10& 0&0&Sat&-&-\\
q2Q10R10-e0i0mN+P2 & 2&10& 10&0& 0&Nep&Nep&-\\
q2Q10R10-e6i0mN-PRY & 2&10& -&0.6& 0&Nep&-&0.01\\
q2Q10R10-e6i0mN-PRO & 2&10& -&0.6& 0&Nep&-&0.001\\
\hline
\end{tabular}
\end{table*}

All planet and asteroid parameters are listed in Table \ref{tab-conf-plt}. These systems of WD, planets and fragments are integrated with the $N$-body package {\small MERCURY} \citep{Chambers1999} for 10 Myr. We have modified the code to implement a simple individual timestepping scheme so fragments very close to the central star that require very small timesteps do not slow down the entire simulation. The details are presented in the Appendix. In the simulation, physical collisions between the fragments and the WD are recorded; those achieving a heliocentric distance of 1000 au are deemed ejected. Additionally, for the simulations with PR-drag, we have also stored the shrinkage time that marks the instant that a fragment's apocentre distance drops below 1 au (then this object is removed from the simulation).

\subsection{Orbits and fate of the tidal fragments}

\begin{figure*}
\includegraphics[width=0.8\textwidth]{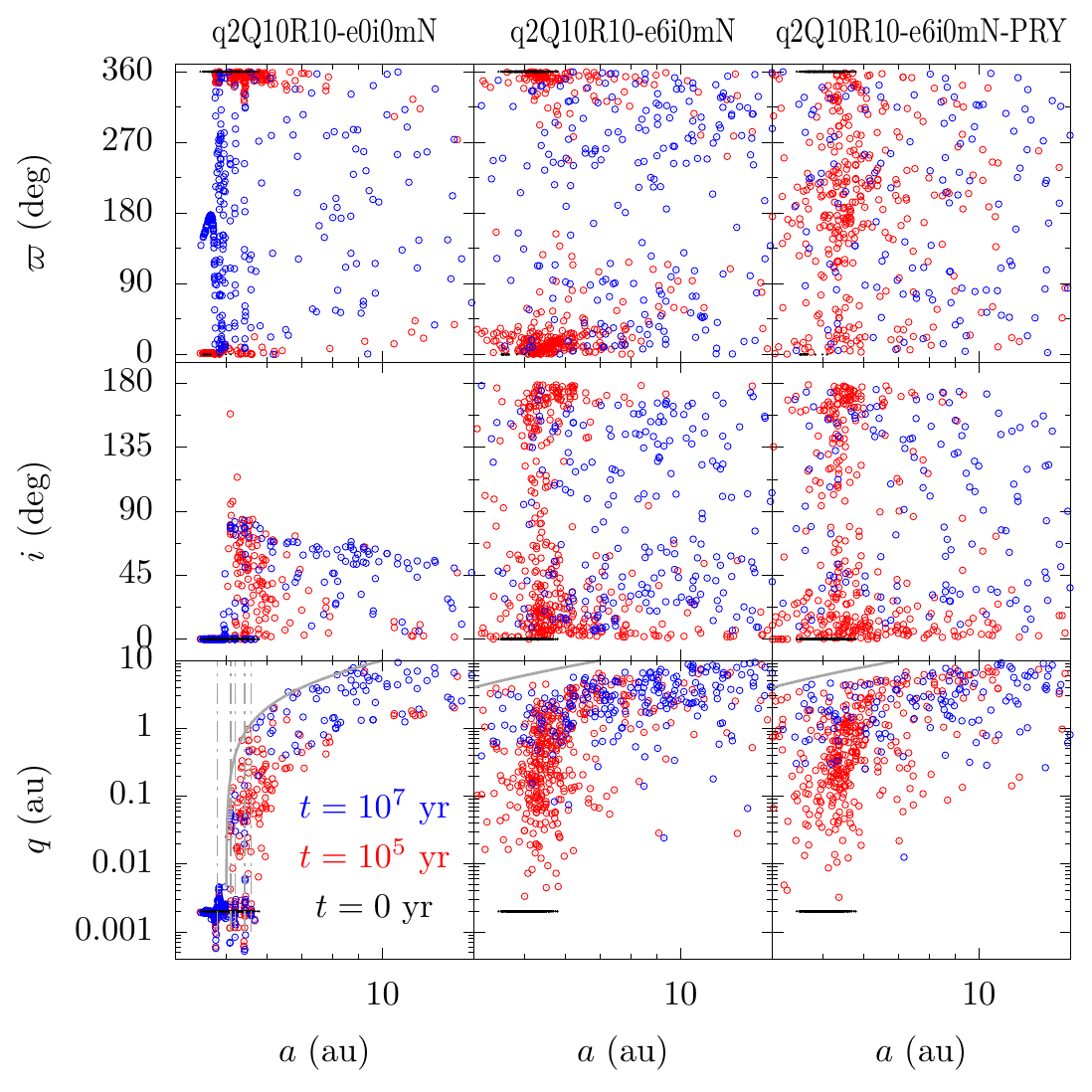}
\caption{The orbital distribution of the surviving fragments at different times from the cases q2Q10R10-e0i0mN (left), q2Q10R10-e6i0mN (middle) and  q2Q10R10-e6i0mN-PRY (right). From top to bottom, the $y$-axes are the longitude of pericentre $\varpi$, inclination $i$ and pericentric distance $q$ and the $x$-axes are always the semimajor axis $a$. The black (small dots), red and blue symbols are measured at 0 yr (immediately after the tidal disruption), $10^5$, and $10^7$ yr, respectively. Also in the bottom-left panels, the grey vertical dash-dotted lines show the locations of some mean motion resonances with the planet (3:1, 11:4, 8:3, 5:2, and 12:5 from left to right); the grey solid line show the direct scattering limit where the apocentre distance of the fragments is equal to the pericentre distance of the planet.}
\label{fig-aeiinfo}
\end{figure*}

First, Figure \ref{fig-aeiinfo} shows the orbital elements of the surviving fragments at different times in q2Q10R10-e0i0mN, q2Q10R10-e6i0mN and q2Q10R10-e6i0mN-PRY from left to right. From top to bottom, the three rows show the distribution of the longitude of pericentre $\varpi$, the inclination $i$ (measured with respect to the planet's orbital plane), and the pericentre distance $q$, all against the semimajor axis $a$.

Clearly visible is that upon tidal disruption (time zero, shown in black), all fragments have the same pericentric distance and inclination as the parent asteroid, forming an vertically thin ring with perfect orbital alignments, consistent with \citet{Debes2012,Veras2014}.% The fact that $\varpi=0^\circ$ merely reflects the way the coordinate system is set. 

The orbits at $10^5$ yr are shown in red. In the left column (q2Q10R10-e0i0mN), fragments with $a>5$ au have been dispersed where their apocentre distances $Q=a(1+e)\sim2a\gtrsim a_\mathrm{P}$. In the bottom panel, $q$ of those outer fragments has been randomised and is just under the scattering limit (the grey solid line). In the meantime, those orbits become highly inclined with a few being retrograde. But the orbits interior to 5 au and decoupled from the planet do not evolve much. In the middle column where the planet eccentricity is 0.6, all fragments are within the reach of direct scattering by the planet and they are all randomised. In the right column where PR-drag is enabled, while the distribution of $i$ and $q$ is similar, there seems to be an overdensity at $\varpi=180^\circ$. Though, PR-drag causes no long-term orbital precession \citep{Veras2015} and we have verified this in test simulations.

Orbits at $10^7$ yr are shown in blue. Now, in addition to more scattering, vertical features in the $(a,q)$ plane of the left column show up. These are where the fragments' mean motions are commensurate with that of the planet and the prominent ratios are 3:1, 11:4, 8:3, 5:2 and 12:5, as shown with the grey dash-dotted lines. These commensurabilities seem to be stable and they can have sizeable libration width at such extreme eccentricity \citep{Wang2017}. Meanwhile, the orbital alignment is now very much broken in that the outer orbits are largely randomised whereas the inner ones show differential precession for q2Q10R10-e0i0mN. The middle and right columns show more extreme orbital randomisation, the main difference being that fragments are sparser in q2Q10R10-e0i0mN+PRY because of removal caused by PR-drag shrinkage.

\begin{figure}
\includegraphics[width=\columnwidth]{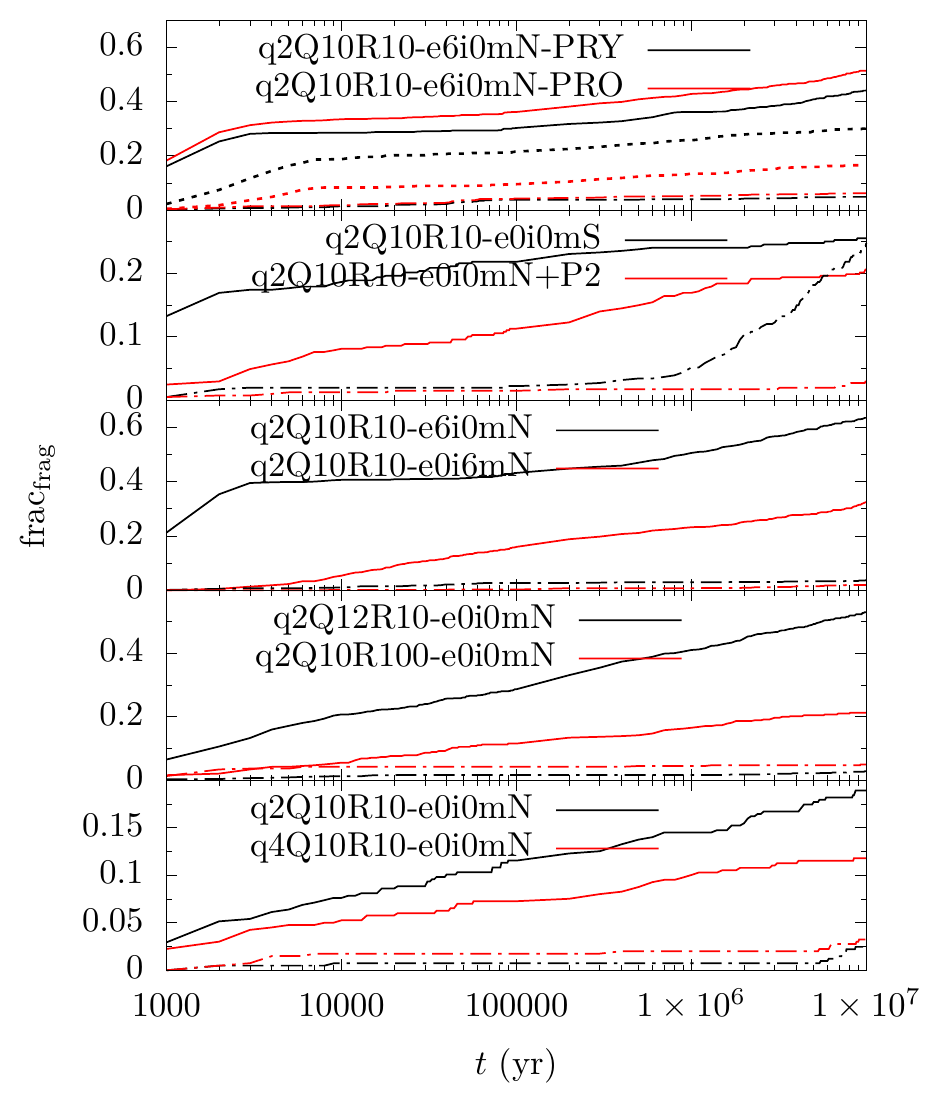}
\caption{Fate of the tidal fragments. Each panel show two planetary system models as the marked with the legend. Solid line show accretion onto the WD, dash-dotted line that of ejection and dotted line that of orbital shrinkage. See Table \ref{tab-conf-plt} for model parameters.}
\label{fig-cdftuns}
\end{figure}

Next, we show the fraction of fragments that have hit the WD, been ejected, or had their orbits shrunk in Table \ref{tab-res}. In general, within 10 Myr, 10\%-50\% of the fragments end up accreted by the WD. The most efficient configuration features an eccentric orbit for the planet \cite[see also][though here the fragments are on WD-grazing orbits already]{Frewen2014} while the least is arguably q4Q10R10-e0i0mN where the initial pericentric distance is twice as large. Also, it seems that PR-drag does not affect the rate of accretion much. When the mass of the planet is small (of Neptunian mass), a few per cent of all the fragments, or less than a tenth of that of consumed by the WD, are ejected. When the planet is as massive as Saturn (q2Q10R10-e0i0mS), ejection and accretion are similarly efficient. Finally, PR-drag shrinks 30\% and 16\% of the fragments' orbits within 10 Myr when the WD luminosity is 0.01 and 0.001 $L_\odot$, respectively.
\begin{table*}
\centering
\caption{Statistics for the tidal fragments at the end of the 10 Myr simulation. The first column shows the simulation label. Shown from the second to the fourth columns are the percentage of fragments that have hit the WD, have been ejected, have their orbits shrunk inside of 1 au. The last column shows the probability that a random pair of fragments has collided with each other. For the set q2Q10R10-e0i6mN, the fragment inclination is measured against the mean orbital plane of all fragments.}
\label{tab-res}
\begin{tabular}{c c c c c }
\hline
label&hit WD (\%)&eject (\%)&shrunk (\%)& coll prob ($\times10^{-7}$)\\
\hline
q2Q10R10-e0i0mN &19&2.5&-&1.1\\
q4Q10R10-e0i0mN &12&3.3&-&1.7\\
q2Q12R10-e0i0mN &53&2.8&-&8.1$\times10^{-2}$ \\
q2Q10R100-e0i0mN &21&5.0&-& 1.1\\
q2Q10R10-e6i0mN &54&3.7&-&1.2$\times10^{-3}$ \\
q2Q10R10-e0i6mN &32&2.0&-&8.0$\times10^{-4}$ \\
q2Q10R10-e0i0mS &26&25&-&4.8$\times10^{-1}$\\
q2Q10R10-e0i0mN+P2 &21&2.9&-&9.3$\times10^{-1}$ \\
q2Q10R10-e6i0mN-PRY &44&4.9&30&- \\
q2Q10R10-e6i0mN-PRO &51&6.3&16&- \\
\hline
\end{tabular}
\end{table*}
{Next, in Figure \ref{fig-cdftuns}, we show the fraction of fragments that hit the WD (solid line) and of those ejected (dash-dotted) and shrunk (dotted) as a function of time. Most models show a similar amount of accretion per decade with no obvious decline (black solid line) except for q2Q10R10-e6i0mN where there is a quick early accretion before $10^4$ yr and perhaps another after 1 Myr. As the top panel shows, PR-drag shrinks the orbits only efficiently within a few times $10^3$ yr}, when the fragments' orbits have not been randomised by the planets and the pericentre distance $q$ is small. Afterwards, $q$ may be perturbed to higher values by the planet.

\begin{figure}
\includegraphics[width=\columnwidth]{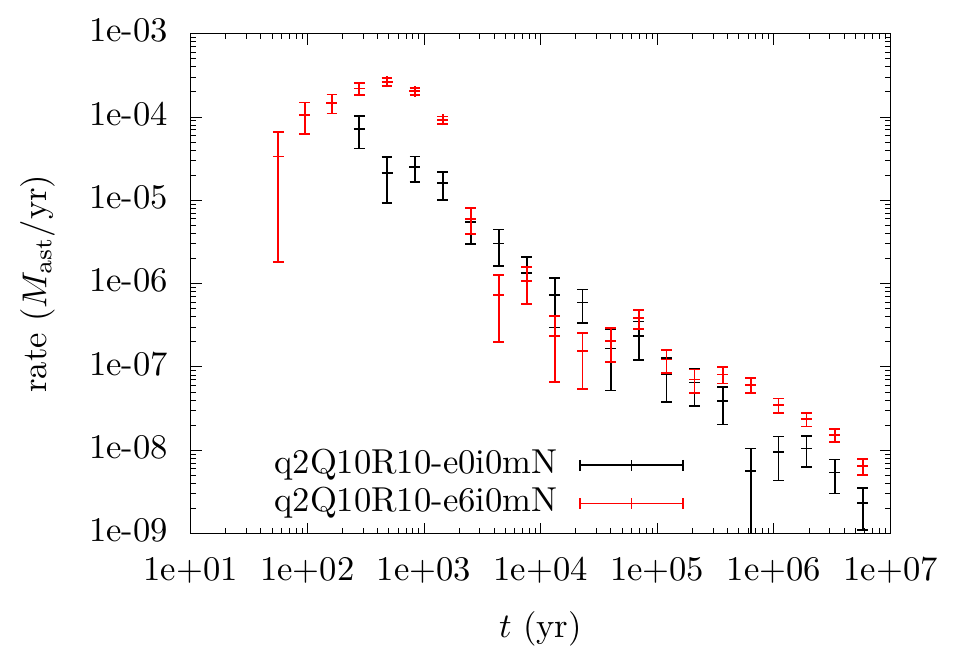}
\caption{The temporal evolution of the accretion rate of fragments in different models. The rate is normalised in the unit of asteroid mass/yr, error bars obtained via bootstrapping.}
\label{fig-acc-rate}
\end{figure}

The fact that tens of per cent of the fragments are accreted indicates a rate of accretion. The rate for two representative models are shown in Figure \ref{fig-acc-rate}. Here for simplicity, we have assumed that each fragment has the same mass and the rate has been measured in the unit of the asteroid mass ($M_\mathrm{ast}$) per year. In the model q2Q10R10-e0i0mN (black), the rate is a few times $10^{-5}$ $M_\mathrm{ast}$/yr at a few thousands of years after the tidal disruption event and steadily drops to $\sim10^{-8}$ $M_\mathrm{ast}$/yr at 1 Myr. This trend is seen in the other models. Though, a few models show interesting short-term behaviour. For instance, the model q2Q10R10-e6i0mN (red) is characterised by an early spike of $\sim10^{-4}$ $M_\mathrm{ast}$/yr at a few thousand years, owing to the direct scattering by the planet before the fragments' orbits are randomised (cf. Figure \ref{fig-praei} below). We note that our accretion rate results from the absorption of the fragments from the tidal disruption of a single asteroid. Various works \citep[e.g.,][]{Farihi2009,Debes2012,Wyatt2014,Petrovich2017,Mustill2018} have derived the rate at which member objects from an asteroid population are delivered onto the WD. For a realistic asteroid population with numerous objects, the time interval between the tidal disruption of two asteroids can be relatively short \citep[for instance][]{Mustill2018}. It is likely that the accretion of the fragments of the preceding asteroid has not finished when the subsequent asteroid is disrupted. In this case, a WD can be actively accreting from multiple asteroids simultaneously.

Not reflected in Table \ref{tab-res} is that PR-drag depends on the fragment size and the WD luminosity. Figure \ref{fig-praei} shows the shrinkage/accretion time of all the fragments as a function of their physical sizes. For q2Q10R10-e6i0mN-PRY ($L=0.01L_\odot$) in the bottom panel, fragments smaller than 1 dm are efficiently shrunk by PR drag on $\sim 10^3$ yr timescale. In q2Q10R10-e6i0mN-PRO ($L=0.001L_\odot$, top panel), the limit becomes 1 cm and in total fewer particles are shrunk. Later shrinkage is less likely as the planet scatters the fragment around and a small pericentre distance may not be maintained. The orbital evolution of larger fragments is dictated by the planets, leading to mainly accretion. Apparent in the plot is that the timing of accretion events shows a gap around $10^4$ yr. Early accretion results from the planet directly scattering the fragments onto the WD whereas for later absorption, the fragments' orbits are first randomised and then scattered onto the WD. The top panel presents the inclination of the particle orbits at the moment of shrinkage, mostly very small especially for those from early shrinkages. This suggests that those still share the same orbital plane, thus disk like when shrunk.

\begin{figure}
\includegraphics[width=\columnwidth]{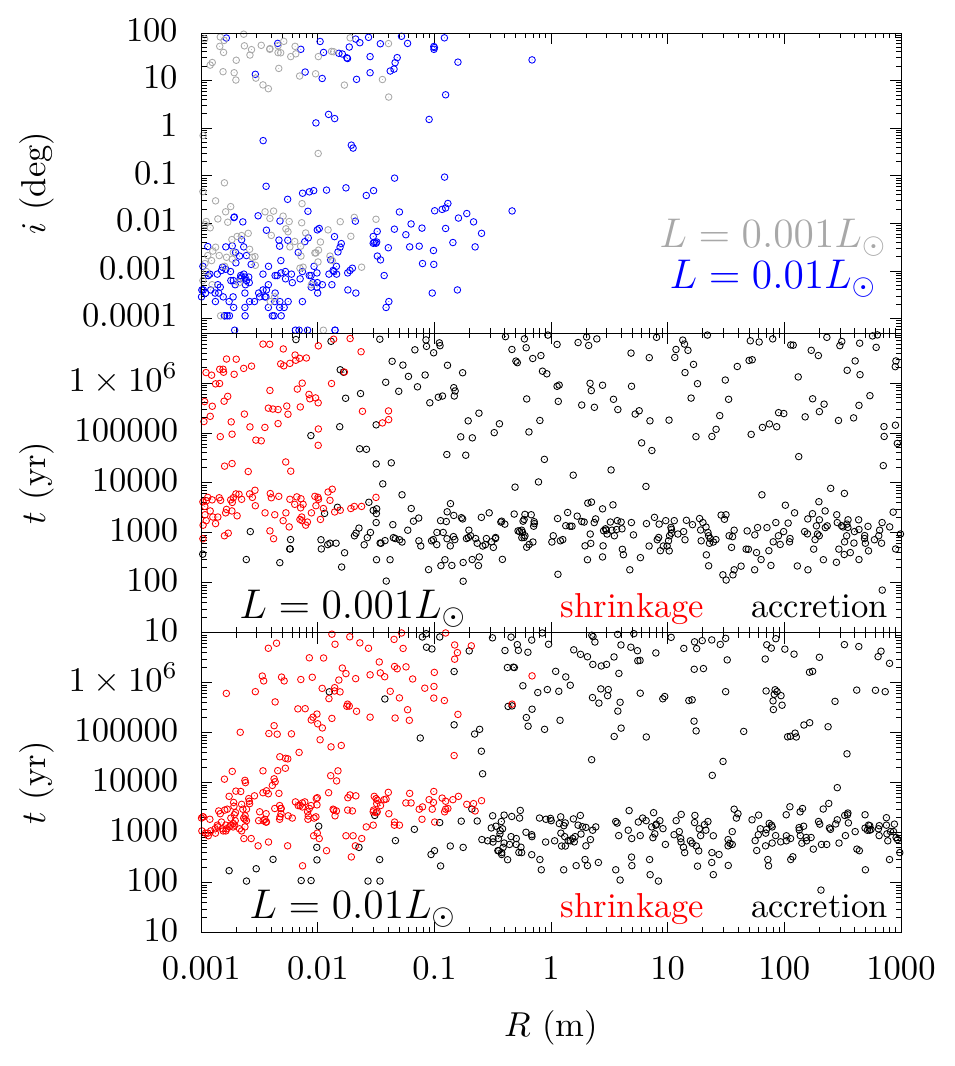}
\caption{Shrinkage, accretion time, and orbital inclination as a function of fragment physical size for simulations where PR-drag is enabled. When shrunk, a fragment's apocentre distance becomes smaller than 1 au. In the bottom two panels, the WD has different luminosity, as labelled in the bottom left corner of each. The top panel shows orbital inclination when shrunk.}
\label{fig-praei}
\end{figure}

Finally, we briefly discuss the implication of the interplay between planet scattering and PR-drag among a real population of tidal fragments. As mentioned in Section \ref{sec-rep} already, monolithic asteroids will be split into fragments of hundreds of metres to a few km, depending on the pericentric distance \citep{Kenyon2017,Rafikov2018,Manser2019}. These large fragments are immune from PR-drag  \citep[perhaps except for WDs at very small cooling ages][]{Veras2015a}. The evolution of them are thus solely determined by planet scattering and from the simulations here, usually leading to accretion in the next few Myr.

If on the other hand, the parent asteroid is a rubble pile, fragment of a wide range of physical sizes from mm to tens of metre result \citep{Mazrouei2014,Michikami2019,DellaGiustina2019}. Then fragments near the lower end (sub-dm) will be affected PR and the simulations show that their orbits will be quickly shrunk and circularised within the Roche lobe of the WD \citep{Veras2015a}. The larger fragments will be accreted onto the WD due to scattering by the planet on Myr timescales. We leave the detailed discussion to Section \ref{sec-con}.% Overall, the fraction of particles are small as shown in Figure \ref{fig-col-rad-ast}.

%%%%%%%%%%%%%%%%%%%%%%%%%%%%%%%%%%%%%%%%%%%%%%%%%
\section{Fragments' mutual collisions}\label{sec-coll}

The mutual interactions between the fragments have so far been omitted. For such small objects with high orbital velocity, as we will see, because of the high relative velocity, mutual collisions can dominate over the gravitational forcing. On short timescales \citep[perhaps over a few orbits][and see Section \ref{sec-rep} of this work]{Debes2012,Veras2014,Malamud2020}, the fragments are collisionless. Here we aim to shed light on the collisional evolution on longer timescales between the fragments.

\subsection{General trend}\label{sec-col-1}
We have modified the {\small MERCURY} code to better detect collisions between small bodies using actual propagation of their state vectors when they are close instead of interpolation. However, test simulations have yielded no detection. Therefore, we resort to an analytical approach by \citet{Kessler1981}. The collision rate between two objects is proportional to the product of the collision cross section $\sigma$, the probability that both two appear at the same location $P$, and their relative velocity $v_\mathrm{col}$. 
\begin{equation}
P_\mathrm{col}\propto \sigma\, P(r,\beta)\, v_\mathrm{col}(r,\beta) \, r^2 \cos\beta,
\end{equation}
where both $P$ and $v_\mathrm{col}$ are a function of the heliocentric distance $r$ and the vertical axis $\beta$. In the derivation of \citet{Kessler1981}, it is assumed that the orbital phases are randomised (but $a$, $e$, and $i$ are kept unchanged) and that the interaction between the two objects can be omitted. The first implies no azimuthal dependence and the second that gravitational focusing is negligible. The collision rate, when integrated over all possible locations of collision (radially and vertically), results in the overall collision probability between the two objects.

\begin{figure}
\includegraphics[width=\columnwidth]{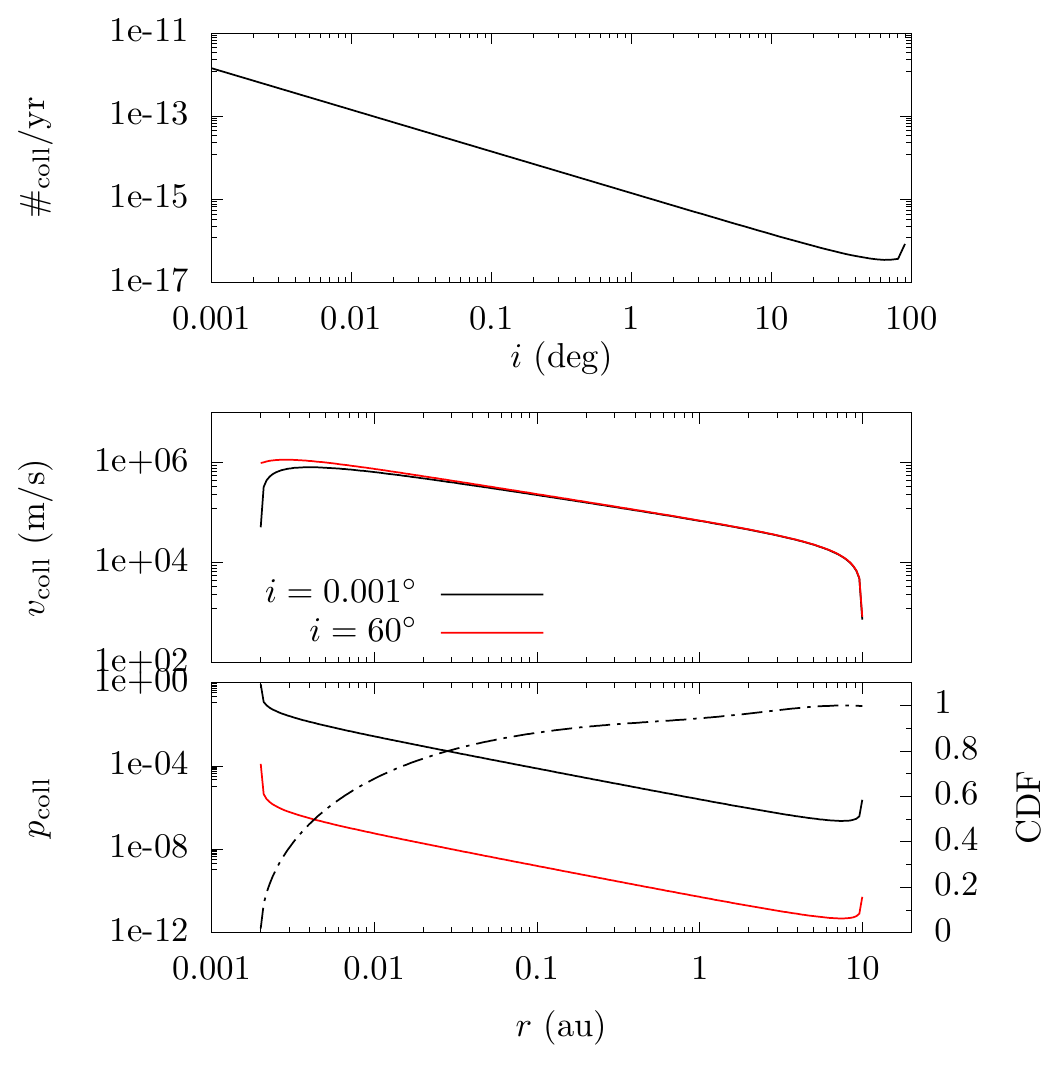}
\caption{Collision probability between two tidal fragments. Both fragments are 1 km in radius, and have the same orbit of pericentre distance of $2\times10^{-3}$ au and apocentre distance of 10 au. In the top panel, the collision rate is shown as a function of the orbital inclination, wherever the collision occurs. In the lower two panels, the collision velocity and rate are plotted in solid lines against the left ordinate as a function of the heliocentric distance of the collision, now fixing inclination at $0.001^\circ$ (black) or $60^\circ$ (red). In the bottom panel, the dash-dotted line shows the normalised cumulative probability function for collision happening inside of $r$, inclination being $0.001^\circ$ against the right $y$-axis.}
\label{fig-q2Q10}
\end{figure}

As a case study, we first consider the collision between two tidal fragments. Both orbits have a pericentre distance of $2\times10^{-3}$ au and an apocentre distance of 10 au; their orbital inclinations are the same and are either $0.001^\circ$ or $60^\circ$. Here we only integrate the collision rate over the vertical direction so the radial dependence can be analysed. The normalised collision probability $P_\mathrm{coll}$ as a function of the heliocentric distance $r$ is shown in the bottom panel of Figure \ref{fig-q2Q10}, $i=0.001^\circ$ in black and $i=60^\circ$ in red. $P_\mathrm{coll}$ attains its maximum at the pericentre and decreases with $r$ but there is a local maximum near the apocentre \citep{Wyatt2010}. The collision probability for $i=60^\circ$ is lower by several orders of magnitude. Additionally, corresponding to the right ordinate, the black dash-dotted line shows the scaled cumulative distribution function, i.e., probability of collision happening inside $r$, for $0.001^\circ$. It is clear that most collisions occur very close to the star \citep{Wyatt2010}, e.g., 70\% inside 0.01 au (less than twice the Roche radius).

The middle panel of Figure \ref{fig-q2Q10} shows the the collision velocity $v_\mathrm{coll}$ for the two inclinations. $v_\mathrm{coll}$ is typically of tens of thousands to hundreds of thousands of m/s and is the largest at small heliocentric distances \citep{Wyatt2010}. At such high collision velocities, gravitational focusing can be safely ignored.

In the top panel of Figure \ref{fig-q2Q10} we have integrated the collision rate both radially and vertically and show the so-obtained number of collisions per Myr between two objects (each assumed 1 km in radius) as a function of the orbital inclination. Overall, collisions are extremely rare and the number decreases rapidly with the inclination and that at $0.001^\circ$ is $10^3$ times that at $1^\circ$. We note that the formalism of \citet{Kessler1981} is singular for a zero inclination, so here the lower limit for inclination is taken $0.001^\circ$ (cf. Figure \ref{fig-praei}). When we work with the actual fragments from Section \ref{sec-coll-frag} in the following, if the inclination of a fragment is smaller than $0.001^\circ$, we let it be $0.001^\circ$.

A potential issue for our application is that, as in Figure \ref{fig-aeiinfo} above, the orbits of the fragments may be actually highly aligned with pericentre pointing toward the same direction, clearly violating the assumption of random orbital phasing of \citet{Kessler1981}. Nonetheless, as we show below, the collision probability so derived is correct to an order of magnitude.

\begin{figure}
\includegraphics[width=\columnwidth]{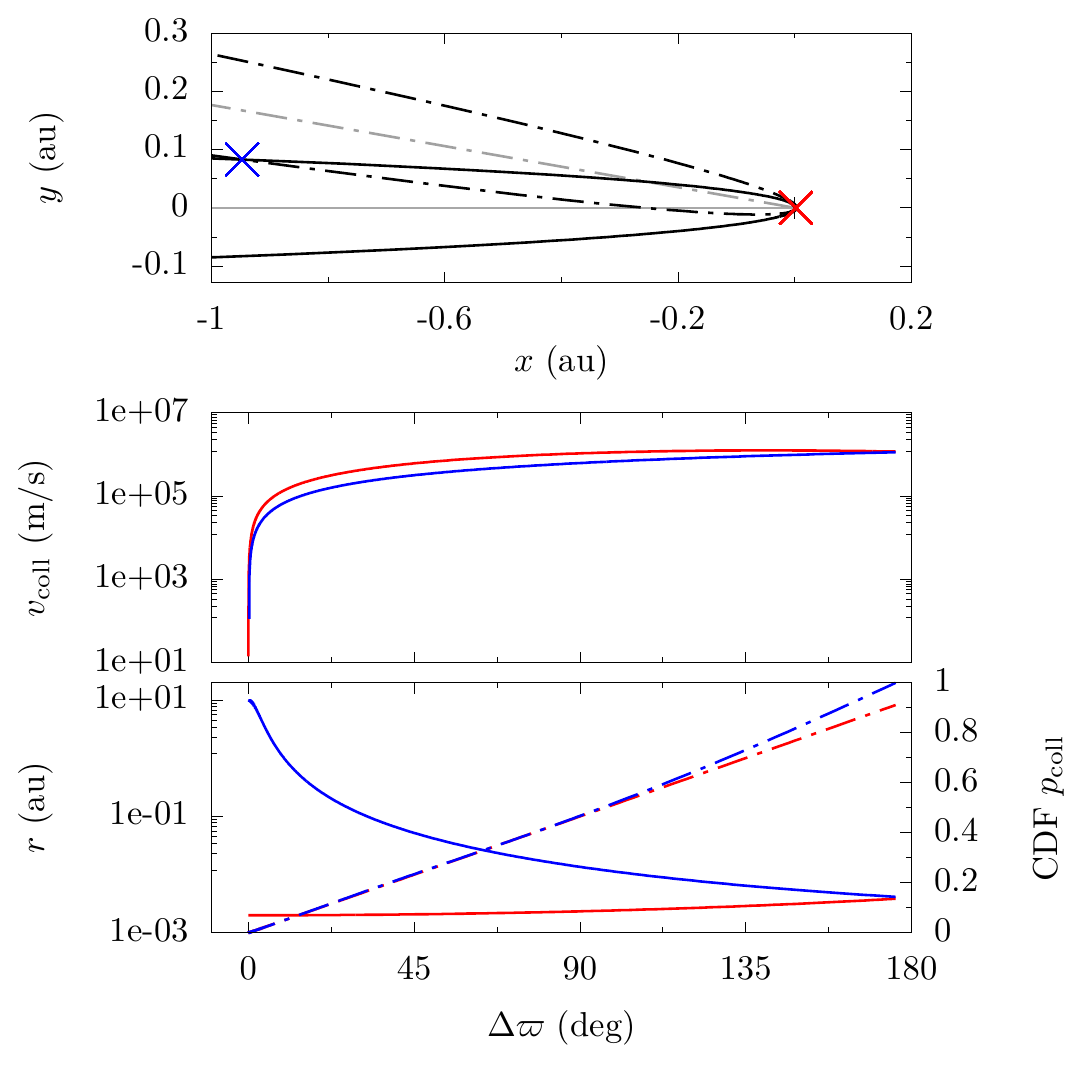}
\caption{Collision configuration between two objects. The orbits of the two are coplanar and both have a pericentre of $2\times10^{-3}$ au and apocentre of 10 au. The directions of the pericentre of the two are offset by $\Delta \varpi$. The top panel illustrates the two possible collision locations with the red (closer) and blue (farther) crosses when $\Delta \varpi=10^\circ$ (the angle between the grey solid and dash-dotted lines). The lower two panels show in solid lines the collision velocity and the heliocentric distance (left $y$-axis) of the closer and farther collisions as a function of $\Delta \varpi$ in red and blue. In the bottom panel, the dash-dotted line shows the relative probability for collisions between two objects with orbital misalignment randomly distributed between $\Delta \varpi\in(0,\Delta \varpi_0)$ (so now the $x$-axis is $\Delta \varpi_0$).}
\label{fig-varpi_r_v}
\end{figure}

In the top panel of Figure \ref{fig-varpi_r_v}, we show an example collision configuration between two objects on coplanar orbits with the same semimajor axis and eccentricity but their orbits are mutually rotated by $\Delta \varpi=10^\circ$. The two orbital intersections, the closer one in red the farther one in blue, dictate where collisions are allowed. The heliocentric distance and the relative velocity of the two are a function of $\Delta \varpi$ as shown in the lower two panels. As $\Delta \varpi$ increases from zero, the closer one of the two collision points starts from the pericentre and very slowly moves outwards while the farther collision point moves in very quickly from the apocentre. The two meet when $\Delta \varpi=180^\circ$. In the meantime, as the two orbits become more and more misaligned, the collision velocity increases and the closer collision is always faster.

The question we want to answer is if the orbital phases of two objects are random but the misalignment is confined to $\Delta \varpi\in(0,\Delta \varpi_0)$, whether or not the value of $\Delta \varpi_0$ affects their collision probability. For any $\Delta \varpi_0$, we obtain the heliocentric distances of the red and blue collisions. Then according to the above reasoning, should $\Delta \varpi\in(0,\Delta \varpi_0)$, closer collisions have to happen interior to the red collision (top panel of Figure \ref{fig-varpi_r_v}) determined by $\Delta \varpi_0$ and farther collisions outside of the blue collision for $\Delta \varpi_0$. So we take the expression for the collision rate from \citet{Kessler1981} and integrate it vertically but radially only in the inner/outer ranges. In this way, the collision probability for $\Delta \varpi\in(0,\Delta \varpi_0)$ is obtained. The normalised probability for the closer and farther collisions is plotted as the red and blue dash-dotted lines in the bottom panel as a function of $\Delta \varpi_0$, corresponding to the right $y$-axis. Both are roughly linear functions of $\Delta \varpi_0$. This means that if we have a given number of objects with orbital misalignment confined to $(0,\Delta \varpi_0)$, their total collisional probability does not rely on $\Delta \varpi_0$. Therefore, the collision probability for $(0^\circ,180^\circ)$ \citep{Kessler1981} is correct in its value even if the fragments' orbits are not fully randomised. In addition, a closer inspection suggests that when the orbits are highly aligned ($\Delta \varpi_0$ is small), red collisions and blue collisions are equally likely. This somehow disagrees with the case of $\Delta \varpi\in(0^\circ,180^\circ)$ where collisions most likely occur close to the star (dashed-dotted line in the bottom panel of Figure \ref{fig-q2Q10}). However, the difference is a factor of two at most.

A similar issue has to do with the ``in-orbit phase''. In the impulse disruption limit, all fragments are launched from the parent asteroid at the same location and at the same time. So their positions on orbit, the true longitude $\varpi+f$ ($f$ is the true anomaly), have been the same at the time of disruption. How fast this quantity disperses can be measured by the filling time, the time needed for the fragments to fill out a ring \citep{Veras2014}. With a spread in $a$ of $\sim$ au, this is only a few tens of yr.

Above we have shown that collisions among the fragments are inclination-dependent. Collisions tend to occur close to the WD (chance of 0.7 inside of 0.01 au) at tens of km/s. Importantly, the analytical method of \citet{Kessler1981} is valid even if the orbital phases of the fragments have not been randomised.

\subsection{Collision among tidal fragments}\label{sec-coll-frag}
\begin{figure}
\includegraphics[width=\columnwidth]{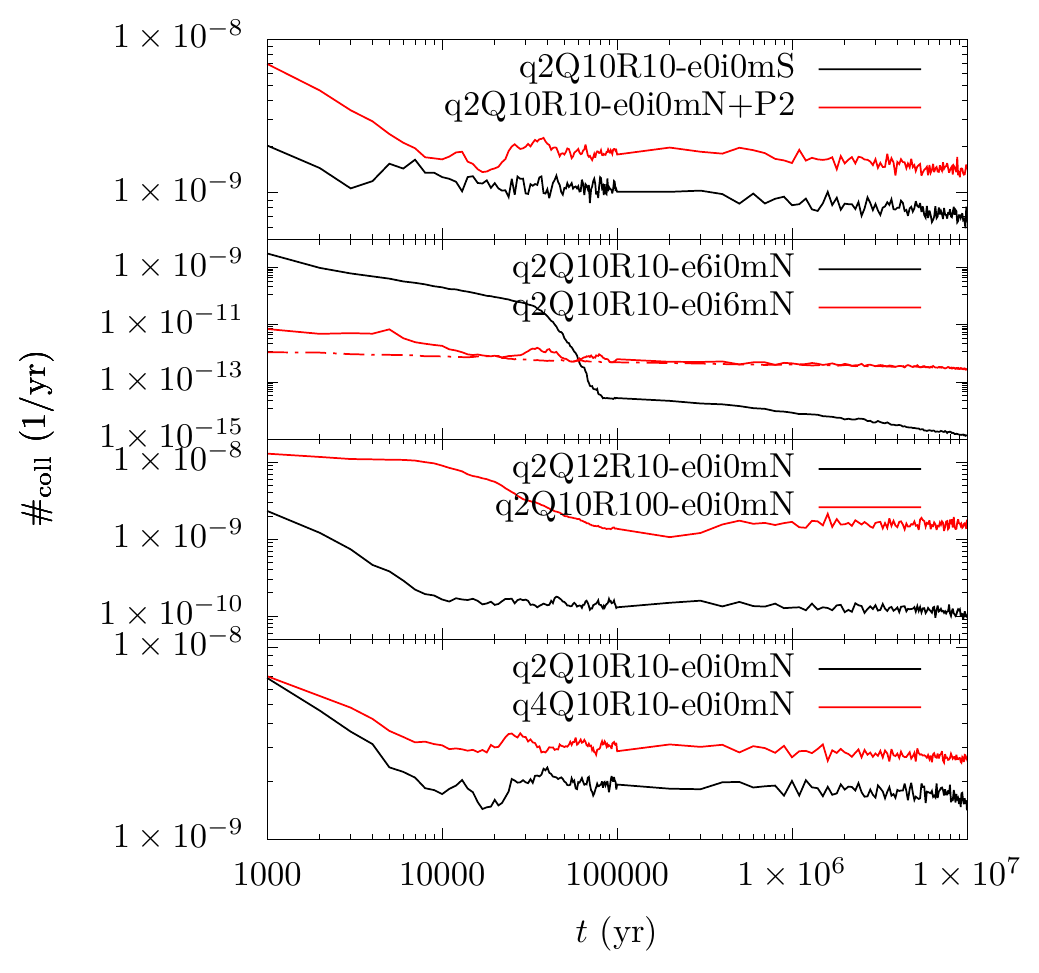}
\caption{Time evolution of collision probability among the fragments in different models. The fragments' orbitals are taken from Section \ref{sec-planet} and fragments physical sizes are all 1 km in radius. In the second panel for the case q2Q10R10-e0i6mN, the red solid line is obtained where the fragments' inclination is measured against the mean orbital plane of all fragments whereas the red dash-dotted line against the planet's orbital plane.}
\label{fig-coll-evo}
\end{figure}

In the following, we use the above approach to calculate the collision probability for the fragment population from the simulations in Section \ref{sec-planet}. For each model in Table \ref{tab-conf-plt} and at every output, the collisional probability between each pair of fragments is analytically calculated using the method of \citet{Kessler1981} where all fragments are assumed to be 1 km in radius. Then these are summed up and a total collision probability is obtained. However, in simulations including PR-drag, random physical sizes have been assigned to the particles and moreover the particles are removed when their orbits are small. This is inconsistent with the equal-size assumption made here so we will not calculate the collision probability for those two sets. The time evolution of the collision probability for the other models is shown in Figure \ref{fig-coll-evo}.

Take q2Q10R10-e0i0mN for instance (black line in the bottom panel). The collision probability is the largest just after the disruption and drops very quickly: $2\times10^{-8}$ collisions per yr among the 400 fragments (at time $t=$ 0 yr and not shown in the logscale plot) within $10^4$ yr timescale, it decreases to $2\times10^{-9}$ because of the early loss of fragments and orbital excitation. Afterwards, the rate largely remains unchanged. The total collision probability is dominated by fragments on low-inclination orbits and those are the inner fragments immune from long-term orbital evolution (Figure \ref{fig-aeiinfo}). We have verified that the overall collision rate corresponds well with the number of surviving fragments on low-inclination orbits.

The evolution of the collision probability in most of the other models follows a similar trend. The one that stands out is q2Q10R10-e6i0mN, where the planet eccentricity is 0.6. Herein, multiple phases of decline are seen. Notably, a sharp plunge kicks in at a few times $4\times10^4$ yr into the simulation and finishes at $8\times10^4$ yr, caused by a quick reduction of near-coplanar fragments.

Another system configuration needing comment is q2Q10R10-e0i6mN. Here arises a question -- in the collision rate calculation, with respect to which plane the fragments' orbital inclination is to be measured. In the long-term, the natural reference plane is the orbital plane of the planet, against which the fragments' orbits precess. Immediately after the tidal disruption, the fragments' orbits, though highly tilted as seen from the planet's ecliptic, reside actually in the same plane and are well aligned. The inclination in the collision rate assessment \citep{Kessler1981} acts to describe the actual vertical spread of the orbits. Therefore, we instead measure the fragments' orbits against the mean orbital plane of all fragments. In such a way, when the orbits lie in the same plane, the coplanarity is naturally captured, and when the orbits have been randomised by the planet, the reference plane will automatically approach the orbital plane of the planet. The collision probability so-obtained is plotted with the red solid line in the second panel of Figure \ref{fig-coll-evo}. We have also performed another calculation where the reference plane is taken to be the orbital plane of the planet and the result is shown as the red dash-dotted line. As expected, the solid line predicts collision probability orders of magnitude higher than the dash-dotted line in the first few thousands yr than the dash-dotted one. But the two quickly converge within $10^4$ yr as the planet-induced orbital precession is fast, with timescales of $10^3-10^4$ yr.

\begin{figure}
\includegraphics[width=\columnwidth]{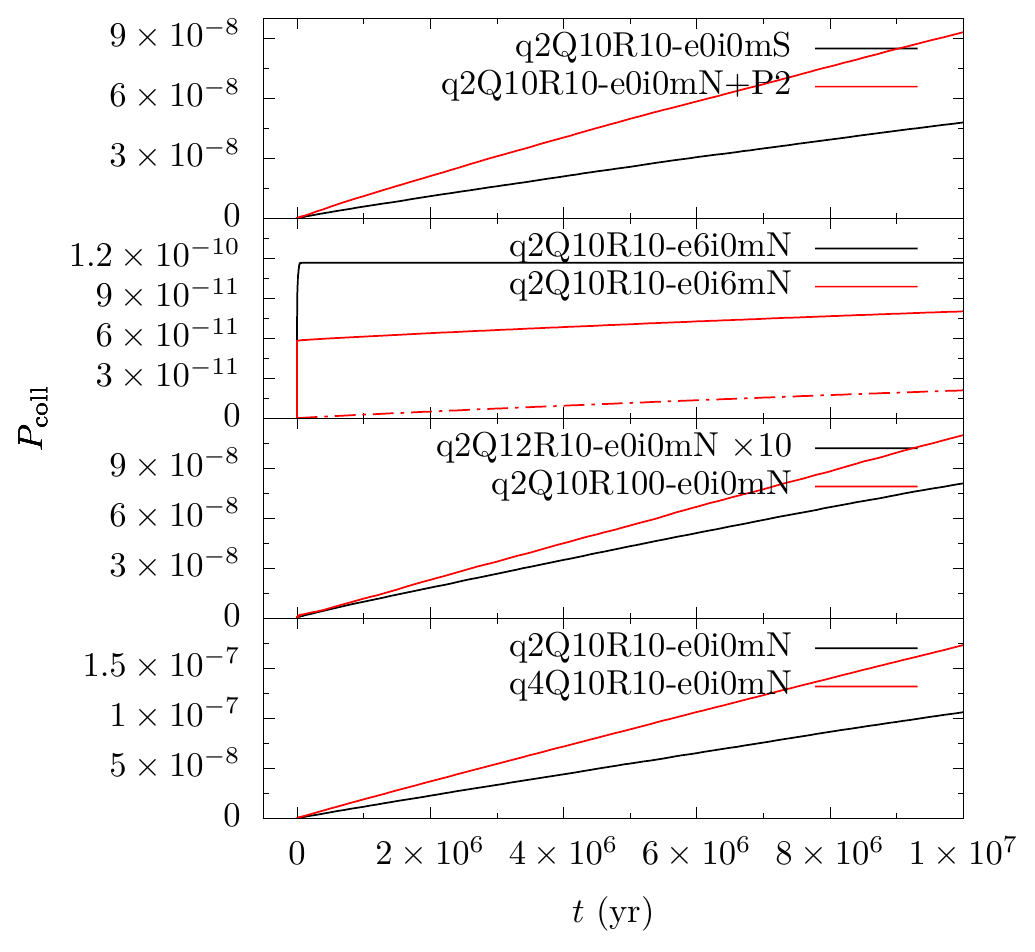}
\caption{The temporal evolution of the time-integrated chance of inter-fragment collision. This quantity shows how likely an average pair of fragments (each 1 km in radius) have collided while not ejected or hitting the WD. In the second panel, the red solid and dash-dotted lines are obtained taking the mean orbital plane of all fragments and the planet’s orbital plane as the reference frame when measuring the inclination, respectively. In the third panel, the chance for q2Q12R10-e0i0mN has multiplied by a factor of 10 for better visibility.}
\label{fig-coll-accu}
\end{figure}

The above collision probability, while capturing the time evolution, is not straightforward when assessing how likely a fragment pair is to have collided. Therefore, we integrate the temporal collision rate above over time and divide the integral by the square of the initial number of fragments. The result can be interpreted as the chance that a random pair of tidal fragments have (survived planet scattering and) collided as a function of time. Table \ref{tab-res} shows the collision probability at 10 Myr and Figure \ref{fig-coll-accu} its time evolution.

We again take the case q2Q10R10-e0i0mN as an example. Over the integration, the chance steadily increases at a constant rate and reaches $\sim10^{-7}$ at 10 Myr. This behaviour is observed for many other cases where the chance at 10 Myr is $\sim10^{-8}-10^{-7}$. 

Two models, q2Q10R10-e6i0mN and q2Q10R10-e0i6mN in the second panel are qualitatively different. The former (black line) is characterised by an initial instant jump and then it levels off immediately, meaning that collisions can only happen very early, before the fragments' orbits are scattered by the planet. For the latter, the reference plane can be the mean of the orbital plane of all fragments (red solid line) or the planet orbital plane (red dash-dotted line). In the former case, the chance of collisions experiences an initial jump followed by a slow growth while in the latter, only the second phase is seen. As argued before, we believe the more proper reference plane is the mean of the orbital planes of all fragments so the case described by the solid red line is more relevant. At 10 Myr, both models predict a collision chance of $\sim10^{-10}$.

The collision chance reported in Figure \ref{fig-coll-accu} varies by orders of magnitude for different models and we need to determine which model is more pertinent to an actual planetary system. Under our assumption that scattering with a planet is directly driving an asteroid towards the WD, probably the asteroid has to experience multiple scattering events and its orbit is likely inclined and intersects with that of the planet \citep{Frewen2014}. Also, if an instability among the planets themselves triggers their scattering with the asteroids \citep{Debes2002,Veras2013,Mustill2018,Maldonado2020}, the planets' orbits are probably eccentric and inclined as well. We deem that the models q2Q10R10-e6i0mN and q2Q10R10-e0i6mN better represent the most realistic scenario. For those two, Section \ref{sec-planet} shows that the planets randomise the fragments' orbits on a timescale of $10^3-10^4$ yr.

\begin{figure}
\includegraphics[width=\columnwidth]{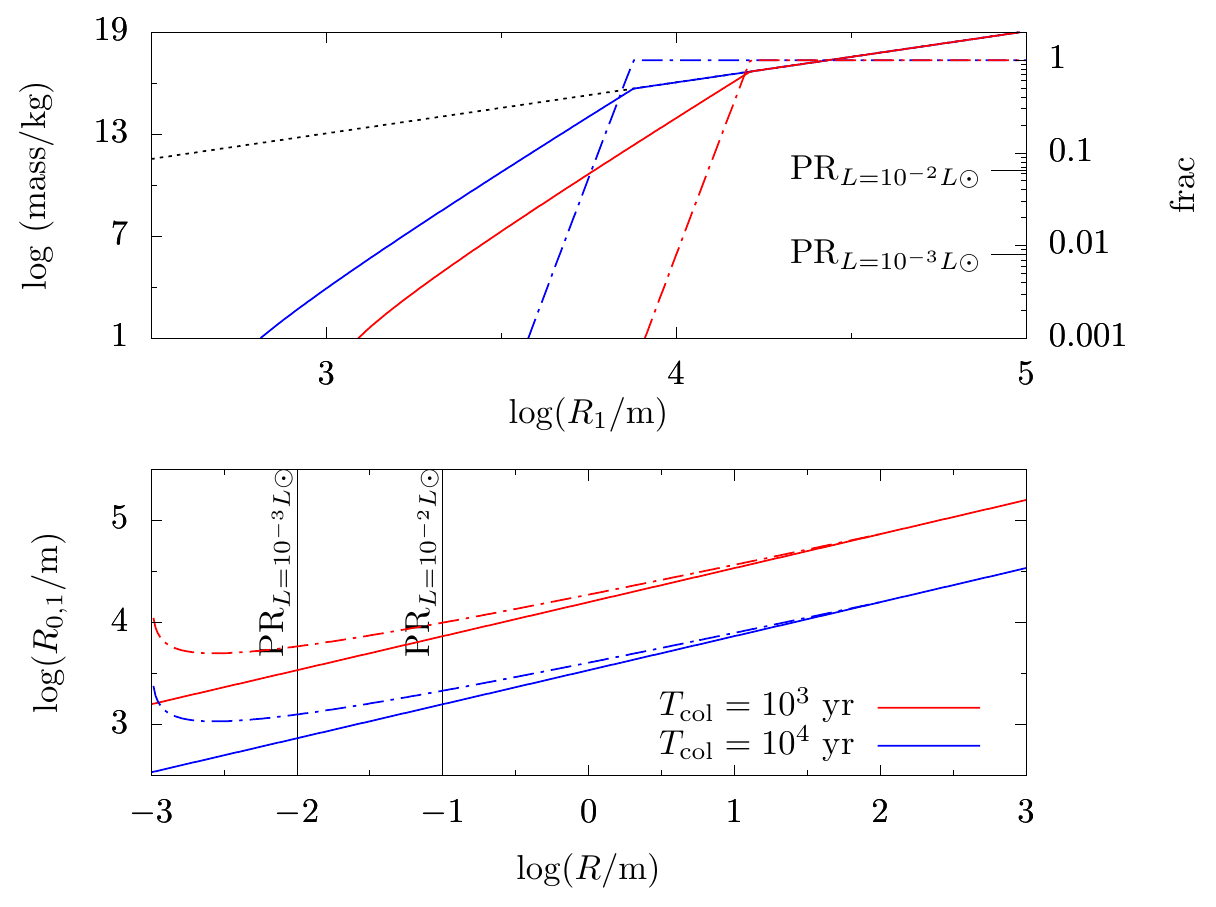}
\caption{Collisions between tidal fragment/particle of an asteroid. In the bottom panel, the $x$-axis is particle size and $y$-axis the asteroid size. For a monodisperse fragment size $R$, the parent asteroid of radius $R_0$ gives rise to just enough number of fragments such that the mutual collision timescale is $10^3$ (red solid line) or $10^4$ (blue solid line) yr. In actuality, because the particle size follows an SFD (e.g., $\in\,$(1 mm,100 m), power law with exponent 3), the asteroid has to be of radius $R_1$ (dash-dotted lines) so particles of $R$ are vulnerable to mutual collisions. The black vertical lines show the limiting $R$ below which PR-drag is effective for two WD luminosities ($0.01L_\odot$ and $0.001L_\odot$). In the top panel, the $x$-axis is $R_1$, left $y$-axis showing the total mass (density 2700 km/m$^3$) of particles lost to mutual collisions (solid lines) and that of the asteroid (black dotted line); the right $y$-axis shows the fractional mass of mutual colliders (dash-dotted lines). The two horizontal lines at the right end shows the mass fraction of small particles subject to PR-drag for the two WD luminosities.}
\label{fig-col-rad-ast}
\end{figure}

Now we discuss the implications of the constraint on the collision timescale. Consider an asteroid of radius $R_0$. Suppose the tidal disruption of it creates $R_0^3/R^3$ fragments, each of radius $R$. For simplicity, all fragments' orbits are assumed the same: $a=5$ au, $q=2\times10^{-3}$ au and $i=0.005^\circ$ (see Figure \ref{fig-praei}). The timescale for a fragment to collide with any of the others can then be estimated using \citet{Kessler1981} as above. There is a one-to-one correspondence between $R_0$ and $R$ for a fixed collision timescale $T_\mathrm{col}$. The solid lines in the bottom panel of Figure \ref{fig-col-rad-ast} show $R_0$ as a function of $R$ for $T_\mathrm{col}=10^3$ (red) and $10^4$ (blue) yr (beyond which mutual collisions become unlikely because of scattering with the planet). Take $R=1$ m for example. In order to have significant mutual collisions between the particles within $10^3$ yr and $10^4$ yr, the parent has to be at least $1.6\times10^4$ m and $3\times10^3$ m in radius, respectively. For small asteroids, simply not enough fragments are generated during the tidal disruption so $T_\mathrm{col}$ is longer. The size of tidal fragments from a monolithic asteroid depends on the material strength and the pericentre distance and are in the range from hundreds of metres to several km \citep{Kenyon2017,Rafikov2018} or perhaps much larger \citep{Manser2019}. Likely the fragments have similar sizes from the same disruption event. Then Figure \ref{fig-col-rad-ast} indicates if fragments are 1 km, only asteroids as large as $1.6\times10^5$ m and $3.4\times10^4$ m can generates large numbers of fragments that experience collision evolution in $10^3$ and $10^4$ yr.

The above assumption of monodisperse fragment size probably does not work for rubble pile asteroids where tidal fragments are single particles composing the parent body. In-situ observations of a few near Earth asteroids have shown that the particles range from mm to tens of m following a power law SFD with exponent close to three \citep{Mazrouei2014,Michikami2019,DellaGiustina2019,Michikami2021}. So here we suppose the particles within a rubble pile asteroid follow an SFD with the power law coefficient of three and that the minimum and maximum particle sizes are 1 mm and 100 m, respectively. 

Now only a fraction of the asteroid is composed of particles smaller than $R$ with large surface area to mass ratios. Suppose that the volume occupied by these $\le R$ particles is $\sim R_0^3$ and that the total volume of the entire asteroid is $R_1^3$. The dash-dotted lines in the bottom panel of Figure \ref{fig-col-rad-ast} show $R_1$ as a function of $R$ for $T_\mathrm{col}=10^3$ (red) and $10^4$ (blue) yr. Apparently, $R_1\ge R_0$. For small $R$, the difference between $R_1$ and $R_0$ is large, meaning that only a small fraction of the total asteroid volume is contained within those small particles susceptible to mutual collision. $R_1>R_0\rightarrow \infty$ when $R=1$ mm (the minimum particle size), and $R_1=R_0$ for $R>$ 100 m (the maximum particle size). Concretely, for $R=1$ m, the parent asteroid is now $1.9\times10^4$ m and $4\times10^3$ m for the two collision timescales.

Inversely, for tidal fragments (or the composing particles) out of a rubble pile asteroid of radius $R_1$, only particles smaller than $R$ with large surface area to mass ratios effectively contribute to the overall collision probability (for $T_\mathrm{col}=10^3$ and $10^4$ yr) and their equivalent total volume is $R_0^3$. Then we assume a density of $2.7\times10^3$ kg/m$^3$ and calculate the mass contained within particles smaller than $R$ as a function of $R_1$, as shown in the top panel with the red and blue solid lines, corresponding to the left ordinate. The black dotted line (left $y$-axis) shows the total mass of the asteroid; and the right $y$-axis (coloured dash-dotted lines) shows the fractional mass affected by mutual collisions. Small asteroids are immune to mutual collisions among the fragment particles. When $R_1<10^4$ m and $<5\times10^3$ m, the mass lost to collision within $T_\mathrm{col}=10^3$ yr and $10^4$ yr only comprises a negligible fraction of the asteroid mass. The transition to the dominance of collision is rather quick -- for asteroids of $1.6\times10^4$ and $<7\times10^3$ m, essentially all asteroid materials suffer from collision in the two cases. The absolute mass that experiences active collisional evolution may vary by orders of magnitude.

\subsection{Synergy between fragment mutual collisions and PR-drag}
As shown in Section \ref{sec-col-1}, the mutual collisions among the tidal fragments occur at relative velocities of tens of km/s or higher \citep[see also][]{Wyatt2010}. Such collisions are probably super-catastrophic, leaving no collisional fragments larger than a tenth of the parent body \citep[e.g.,][]{Benz1999,Leinhardt2012}. These second generation fragments may be subject to further mutual collision.

Among a population of $N$ fragments, the timescale for one to collide with any of the other is inversely proportional to the total collision cross section, or equivalently, proportional to $\propto N^{-1} r^{-2}$ ($r$ is the radius of a fragment). After one collision, consider a simple case where the fragments are split into ten equal mass pieces. The collision timescale between these second generation fragments would be $10^{-1} (0.1^{1/3})^{-2}\approx 0.5$ times that among the progenitor bodies. Hence, the mutual collisions among the collisional fragments have exponentially shorter and shorter timescales -- the fragments will be ground down to tiny dust particles if the timescale of the first generation fragments (the tidal fragments) is shorter than the planet scattering timescale.

Here in the discussion above, we have only addressed cases where PR-drag is inactivated. But what if this effect is at work? As shown in that Section \ref{sec-planet} \citep[and see also][]{Veras2015a}, PR-drag serves to shrink the fragments' orbits but is only efficient for small objects. While large tidal fragments are not affected directly, if mutual collisions are triggered, their collisional fragments may be prone to this effect. As these second-generation fragments' orbits contract, their mutual collision timescale also decreases because of the faster orbital revolution. This may further accelerate the collision process, in addition to the increase of collision cross section as discussed above. Hence, it seems that fragments' mutual collisions tend to occur in a run away manner, perhaps leading to the formation of a dust disk \citep{Veras2015a}.

\section{Discussion}\label{sec-dis}
We have shown that when scattered close to a WD, an asteroid will tidally disrupted and then depending on the asteroid size and internal structure, the fragments may be either directly accreted by the WD by scattering with a planet or processed into a close-disk before accretion. How does direct accretion of the fragments from an asteroid due to sporadic scattering by the planet compare with accretion from a disk formed from an asteroid? While the accretion timescale of the former is as shown here several Myr, accretion from a disk, perhaps through PR-drag and aerodynamic drag, is much quicker of the order of $10^5$ yr \citep{Rafikov2011,Rafikov2011a}. In addition, as we have shown, only massive asteroids can lead to the formation of disks. Then accretion from a disk formed out of a (massive) asteroid is higher than that directly from an (average) asteroid, not inconsistent with the fact that faster accreting WDs have a higher disk occurrence rate \citep{Farihi2009}. We note that here we have only discussed the isolated accretion from a single body whereas in a population of asteroids, another object could enter the Roche radius before a previous one is accreted (in many My) \citep{Mustill2018} and a disk may be replenished by traversing objects \citep{Grishin2019a,Malamud2021,Rozner2021}.

While a few tens of per cent of the WDs are actively accreting asteroidal material, only a tenth of those accreting have dust disks \citep[e.g.,][]{Farihi2009}. We have shown that only large enough asteroids can initiate mutual collisions between the tidal fragments and lead to the formation of a disk before planet scattering dominates. In the solar system, asteroids $\lesssim10$ km are probably rubble piles \citep[e.g.,][]{Bottke2005,Walsh2018}, but those are too small to trigger fragments' mutual collisions. Larger asteroids are likely monolithic. But for those, a larger radius $\sim 100$ km is required for the formation of a dust disk. Hence, it seems that in the solar system, whether of rubble pile or monolithic nature, tidal disruption of asteroids $\gtrsim100$ km will lead to a dust disk and otherwise not. The main belt asteroids are highly top heavy: only a few tens of them are larger than the 100 km threshold. Combined, this suggests that the rarity of disk comes from the rarity of large asteroids \citep[and the short lifetime of a disk][]{Rafikov2011}.

Notably, it seems that for WDs with a dust disk, as indicated by an infrared excess, infrared variation is often observed \citep[see, for example][and references therein]{Swan2020}. For instance, \citet{Wang2019} reported an infrared outburst which they attributed to the tidal disruption of an asteroid. As shown here \citep[also][]{Veras2014}, it takes multiple orbits for the tidal fragments to spread to fill the entire orbit. In the meantime, as the objects that are passing by the pericentre are becoming more numerous and their contribution may lead to an increase in the infrared luminosity. In this process, the fragments' mutual collisions, occurring at close heliocentric distance \citep[see  Figure \ref{fig-q2Q10} and ][]{Wyatt2010}, would probably also create chunks of dust, enhancing the infrared excess further. Therefore, we are more likely to see one of the dozens of subsequent pericentre passages than the original tidal disruption event.

Diffusion timescales for metal sinking in hot DAs are extremely short \citep{Koester2014}. This suggests that maintaining substantial metal in the WD atmosphere needs continuously in-falling asteroids. \citet{Mustill2018} found that delivery of asteroids to the WDs with cooling ages $\lesssim100$ Myr was infrequent. The fast decline of the accretion rate of the tidal fragments from an asteroid on $\ll$ Myr timescales (Figure \ref{fig-acc-rate}) corroborates the above result. On much longer timescales, the change of accretion rate is dictated by the frequency at which the asteroids are scattered onto the WD \citep[e.g.][]{Bonsor2011,Debes2012,Petrovich2017,Mustill2018}.

Lastly, transiting circum-WD planetesimals have been detected in a few systems. That of ZTF J013906.17+524536.89 has a period of about 100 days, suggesting an apocentric distance of 0.7 au \citep{Vanderbosch2020}, possibly a system caught in the process of orbital shrinkage. The WD 1145+017 and ZTF J0328-1219 systems have transiting material on $\lesssim$ 10-hr orbits, inside/close to the Roche radius \citep{Vanderburg2015,Vanderbosch2021}. However, in our model where PR drag is sole mechanism able to contract the fragments' orbits and works only for tiny fragments, major bodies able to sublimate dust clouds persistently are not expected close to the WD. However, those large objects may be brought in through other mechanisms, for instance, disk--asteroid interaction or tides \citep{OConnor2020,Malamud2021,Rozner2021,Zhang2021}.

In multiple-planet systems, depending on the architecture, a few to tens of per cent of unstable planets can be driven to pass within the WD Roche radius \citep{Debes2002,Veras2013,Veras2016a,Maldonado2020,Maldonado2020a,Maldonado2021}. The tidal disruption of these large objects should lead to the formation of a disk as well.

We have not considered the production of gas. Mutual fragment collisions occur at velocity of tens of km/s where tens of per cent of the colliders may vapourise \citep{Kenyon2017,Malamud2021}. Moreover, WD radiation may sublimate the dust particles directly \citep{Rafikov2011,Rafikov2011a,Steckloff2021}. Otherwise, gas could be released from an asteroid made of strong material within the Roche radius \citep{Trevascus2021}. We refer to \citet{Malamud2021} for an extended discussion on this matter.

Here we have only considered the synergy between WD radiation and the mutual collisions between the tidal fragments. Other orbital shrinkage mechanisms, \citep[e.g.,][]{Grishin2019a,OConnor2020,Malamud2021,Zhang2021} that are size dependent may also benefit from the collisional grinding; see \citet{Rozner2021} for a discussion. Similarly, the planet scattering timescale sets a limit to all these effects.

In our scenario, it is the direct scattering between the tidal fragments and a planet that puts a limit on the collisional evolution of the former. However, if the asteroid is pushed towards the WD's Roche radius through secular forcing \citep[for instance][]{Kratter2012,Stephan2017,Petrovich2017}, perhaps such a timescale constraint would be greatly relieved. Then, the fragments would have more time to interact with one another and the chance of disk production could be increased.

\section{conclusion}\label{sec-con}
We have studied the tidal disruption of an asteroid by a WD in its Roche lobe and the ensuing evolution of the fragments. Modelled as a rubble pile and on an extremely eccentric orbit, the asteroid is shredded by the WD tidal field into its constituent particles after a few pericentric passages, resulting in a flat and aligned ring of particles. In assessing the fragment particles' long term evolution, we have considered the perturbation from the planet that is responsible for scattering the asteroid close to the WD, radiation forces (PR-drag) by the WD and the fragments' mutual collisions. We find that depending on the planet's orbit and mass, it scatters the fragments' orbits on timescale of $10^3-10^4$ yr, breaking the alignment and coplanarity. Scattered around the system, tens of per cent of the fragments are accreted by the WD over a few Myr. Radiation forces work on small sub-cm or -dm particles only and shrink their orbits efficiently within $10^3-10^4$ yr before the planet (stochastically) raises the pericentre distance. Active mutual fragment collisions depend crucially on coplanarity and are hence only relevant before the fragments' orbits are scattered by the planet. Assuming that an asteroid is split into equal-size fragments, we show that only large enough parent asteroids can create enough fragments such that collisions are important within $10^3-10^4$ yr. For example, if the fragment size is 1 m, the asteroid has to be $>3$ km in radius and larger fragment sizes require larger asteroids. Then we have tested an SFD typical of near Earth asteroids and find that only for asteroids $>10$ km, a substantial fraction of the asteroids' mass is subject to mutual collisions. At tens of km/s or more, the mutual collisions may effectively grind down the tidal fragments and once the collisional fragments are small enough, PR-drag kicks in and a circumstellar disk results.

Based on the above result, we propose that the tidal disruption of an asteroid around a WD and the long term evolution of the fragments depends on the asteroid properties.
\begin{itemize}
\item During the tidal disruption, monolithic asteroids are broken down into pieces that can withstand the WD tide. Depending on the material strength, the resulting fragments are km sized or much larger \citep{Kenyon2017,Rafikov2018,Manser2019}. These large objects have small surface area to mass ratios and are not affected by radiation forces \citep[][and see Figure \ref{fig-praei}]{Veras2015a}. Whether mutual collisions can come into play relies on the parent asteroid size.
\begin{itemize}
\item If the asteroid is small, $\lesssim 100$ km in radius (bottom panel of Figure \ref{fig-col-rad-ast}), simply not enough fragments are created and the collision timescale is longer than a few $10^3-10^4$ yr. The fragments' long term evolution will be dominated by the interaction with the planet that scatters the parent asteroid to the WD and over Myr timescale, a few tens of per cent of them end up accreted by the WD (Table \ref{tab-res}).
\item If the asteroid is large, $\gtrsim 100$ km in radius, high speed disruptive mutual collisions among the tidal fragments are expected to happen within $10^3-10^4$ yr. Then, collisions among these collisional fragments will occur on shorter and shorter timescales, generating large amount of small dust particles. The orbits of these small objects, under the effect of PR-drag, shrink, leading to a dust disk inside the WD Roche lobe \citep{Veras2015a}.
\end{itemize}
\item Rubble pile asteroids are shredded into the constituent particles during the tidal disruption. Those particles cover a large range from mm to tens of m \citep{Mazrouei2014,Michikami2019,DellaGiustina2019}. The smallest particles are directly affected by PR-drag but only constitute $<10\%$ of the asteroid total mass (top panel of Figure \ref{fig-col-rad-ast}). Whether particle mutual collisions are important within $10^3-10^4$ yr again depends on the asteroid size.
\begin{itemize}
\item If the asteroid is small, $\lesssim 10$ km in radius (top panel of Figure \ref{fig-col-rad-ast}), a small fraction of the asteroid mass contained with small particles are subject to mutual collisions, later converted into a dust disk with the help of PR-drag. The majority of the asteroid mass will then be controlled by the scattering with the planet.
\item If the asteroid is large, $\gtrsim 10$ km in radius, all its constituent particles participate in the mutual collisions and the formation of a disk.
\end{itemize}
\end{itemize}

In summary, our work shows that not all accretion of asteroidal material onto WDs takes place through a disk, the occurrence rate of which depends on the asteroid's size and internal properties.

\section*{Acknowledgements}

The authors thank the anonymous referee for the insightful feedback. The authors acknowledge financial support from Knut and Alice Wallenberg Foundation (2014.0017 and 2012.0150), from Vetenskapsrådet (2017-04945), and from the Royal Physiographic Society of Lund (F 2019/769). Computations were carried out at the center for scientific and technical computing at Lund University (LUNARC).
%%%%%%%%%%%%%%%%%%%%%%%%%%%%%%%%%%%%%%%%%%%%%%%%%%

\section*{Data Availability}
The data underlying this article will be shared on reasonable request to the corresponding author.


%\bibliographystyle{mnras}


%%%%%%%%%%%%%%%%%%%%%%%%%%%%%%%%%%%%%%%%%%%%%%%%%%

%%%%%%%%%%%%%%%%% APPENDICES %%%%%%%%%%%%%%%%%%%%%

\appendix
\section{Individual time-stepping}
The original {\small MERCURY} package uses a global timestep. This timestep is limited by the objects that are close to the central star, potentially slowing down the simulation. In order to alleviate this issue, we simply introduce subsystems containing only fragments close to the WD. For any object achieving a stellar-centric distance smaller than a critical value $d_\mathrm{crit}$, a subsystem is spawned. Herein, only the star, the planets \citep[big bodies as in {\small MERCURY}][]{Chambers1999} and this fragment (as a small body) are considered; this subsystem is propagated in isolation and until the heliocentric distance of the fragment is large again, it collides with the central star, or an output is due. Then the fragment is passed back to the main simulation for further integration, logging the collision or normal output. In the last scenario, the fragment continues to be simulated in a subsystem after the output. When multiple fragments are close to the star, more than one subsystems are created and are dealt with simultaneously. The Bulirsch-Stoer integrator \citep{Press1992} is used for the simulation of the subsystems. Test simulations show that, for a system of a few hundreds of fragments, our revision offers a speedup of a factor of a few for $d_\mathrm{crit}$ around a few time 0.1 au to 1 au. In the simulation in the main body of the paper, we let $d_\mathrm{crit}=0.5$ au.

% Don't change these lines
\bsp	% typesetting comment
\label{lastpage}
\end{document}